\documentclass[proceedings]{JHEP3}
\usepackage{amsfonts}
\usepackage{amsmath}
\usepackage{epsfig,multicol}

\newbox\mybox

\newcommand\fverb{\setbox\mybox=\hbox\bgroup\verb}
\newcommand\fverbdo{\egroup\medskip\noindent\fbox{\unhbox\mybox}\ }
\newcommand\fverbit{\egroup\item[\fbox{\unhbox\mybox}]}
\conference{PT-symmetrically deformed shock waves}
\abstract{We investigate for a large class of nonlinear wave equations, which allow for shock wave formations, how these solutions behave when they are $\mathcal{PT}$-symmetrically deformed. For real solutions we find that they are transformed into peaked solutions with a discontinuity in the first derivative instead. The systems we investigate include the $\mathcal{PT}$-symmetrically deformed inviscid Burgers equation recently studied by Bender and Feinberg, for which we show that it does not develop any shocks, but peaks instead. In this case we exploit the rare fact that the $\mathcal{PT}$-deformation can be provided by an explicit map found by Curtright and Fairlie together with the property that the undeformed equation can be solved by the method of characteristics. We generalise the map and observe this type of behaviour for all integer values of the deformation parameter $\varepsilon$. The peaks are formed as a result of mapping the multi-valued self-avoiding shock profile to a multi-valued self-crossing function by means of the $\mathcal{PT}$-deformation. For some deformation parameters we also investigate the deformation of complex solutions and demonstrate that in this case the deformation mechanism leads to discontinuties.}

\title{$\mathcal{PT}$-symmetrically deformed shock waves}
\author{Andrea Cavaglia and Andreas Fring \\
Centre for Mathematical Science, City University London,\\
Northampton Square, London EC1V 0HB, UK\\
E-mail: andrea.cavaglia.1@city.ac.uk, a.fring@city.ac.uk}

\input{tcilatex}

\begin{document}

\section{Introduction}

Since the proposal that complex $\mathcal{PT}$-symmetric quantum mechanical
Hamiltonians may be viewed as self-consistent descriptions of physical
systems \cite{Bender:1998ke}, $\mathcal{PT}$-symmetry has been exploited to
propose and study many more complex extensions of real systems. Numerous new
quantum mechanical and quantum field theoretical models have been
investigated, see for instance \cite{Benderrev,Alirev,PauloPhD} for recent
reviews. Inspired by this success, the construction principle has also been
used to suggest new classical models, such as complex extensions of standard
one particle real quantum mechanical potentials \cite%
{Nana,Bender:2006tz,Bender:2008fr,Bender:2009jg,Bender:2010eg,Benderbands},
non-Hamiltonian dynamical systems \cite{Bender:2007pr}, chaotic systems \cite%
{Bender:2008qe} and deformations of many-particle systems such as
Calogero-Moser-Sutherland models \cite%
{AF,AFZ,Assis:2009gt,FringSmith,Fring:2011fm,Ghosh1,Ghosh2}. Here we will
mainly focus on extensions of nonlinear wave type, such as the prototype
Korteweg-deVries (KdV) equation \cite{BBCF,AFKdV,BBAF} and closely related
models \cite%
{Bender:2007ij,Curtright:2007ta,comp1,CompactonsAF,Cavaglia:2011qh}. In \cite%
{Cavaglia:2011qh} it was demonstrated that when reduced these sytems can
also shed light on large classes of one particle quantum mechanical models.

As is well known from transformation theory a lot of new information on
nonlinear wave equations can be obtained from transformed equations by means
of for instance Hopf-Cole, Miura or B\"{a}cklund type. In a somewhat similar
spirit we also exploit here the knowledge of a transformation in form of an
explitly known $\mathcal{PT}$-symmetrical deformation. Here we will mainly
focus on the question of what kind of effect a $\mathcal{PT}$-deformation
has on a shock wave. It is well known that a shock forms when the crest of a
wave overtakes the troughs. The challenge for a mathematical description is
that one can no longer describe this phenomenon by a function since the
surface of the wave becomes multi-valued. For real wave equations solvable
with the method of characteristics, this happens when two characteristics
cross each other or more generally when the first derivative becomes
infinite. For $\mathcal{PT}$-deformed equations remaining real, we argue
here that the first derivative is discontinuous but remains finite, whereas
the second derivative tends to infinity. Solitonic solutions with this type
of behaviour are often referred to as peakons \cite{Peakon}. When the
deformation parameters are non odd integers, one is forced to consider
complex solutions even in the undeformed case if one demands a real
soltution for the deformed one. However, evolving this real deformed
solution in time will convert it into a complex one. In addition, when
compared to the real scenario, the peaks vanish and we observe
discontinuties, which are generated due to the imposition of physical
asymptotic boundary conditions.

Our manuscript is organised as follows: In section 2 we describe the $%
\mathcal{PT}$-deformed models we are investigating and how the explicit
knowledge of the deformation map can be utilised to extract information
about the systems, in particular the shock time and conservation laws. In
section 3 we describe the general mechanism of how real shock waves are
mapped into peaks and in section 4 how a modification of this mechanism
leads to jumps in a complex scenario. In section 5 we present various
numerical case studies supporting and illustrating our findings. We present
our conclusions in section 6.

\section{Shock times and conservations laws from $\mathcal{PT}$-deformation
maps}

The model we wish to study here with regard to shock wave and peak formation
is a $\mathcal{PT}$-symmetrical deformation of a nonlinear wave equation 
\begin{equation}
\mathcal{PT}_{\varepsilon }:\qquad w_{t}+f(w)w_{x}=0\quad \rightarrow \quad
u_{t}-if(u)(iu_{x})^{\varepsilon }=0,  \label{InvB}
\end{equation}%
with $f(w)$ being a well behaved function of $w$. Clearly the undeformed
equation is $\mathcal{PT}$-symmetric, that is being invariant under a
simultaneous reflection in time $t\rightarrow -t$ and space $x\rightarrow -x$%
, when $\mathcal{PT}:f(w)\rightarrow f(w)$, $w\rightarrow cw$ with $c\in 
\mathbb{C}$. The deformed equation is constructed as usual by taking into
account that the $\mathcal{PT}$-transformation is antilinear \cite{EW} and
an overall minus sign in the second term is generated from $i\rightarrow -i$
rather than from $x\rightarrow -x$. The special case $f(w)=w$ corresponds to
the $\mathcal{PT}$-symmetrical deformation of the inviscid Burgers equation,
also referred to as Riemann-Hopf or Euler-Monge equation, recently studied
by Bender, Feinberg \cite{Bender:2007ij} and Curtright, Fairlie \cite%
{Curtright:2007ta}. The real version of the first equation in (\ref{InvB}),
especially for $f(w)=w$, appears mainly in fluid mechanics whereas its
complex version is frequently encountered in the treatment of large N matrix
models, see e.g. \cite{Gopakumar:1994iq}.

Most wave equations are only directly related to their $\mathcal{PT}$%
-symmetric deformations in the limit $\varepsilon \rightarrow 1$ and an
explicit transformation between the two is not known otherwise. For
instance, this is the case for the deformations of the KdV-equation \cite%
{BBCF,AFKdV,BBAF,Cavaglia:2011qh}. In contrast, in some rare cases the
deformation map $\mathcal{PT}_{\varepsilon }$ constitutes an explicit
transformation from the undeformed to the deformed system. As pointed out by
Curtright and Fairlie \cite{Curtright:2007ta} for the system (\ref{InvB})
with $f(w)=w$, the first equation in (\ref{InvB}) converts into the second
under the map $w\mapsto \varepsilon u(iu_{x})^{\varepsilon -1}$. We
generalise this here to equations with arbitrary $f(w)$ to 
\begin{eqnarray}
\mathcal{PT}_{\varepsilon }^{\prime }:\qquad &&w\mapsto \varepsilon
f(u)(iu_{x})^{\varepsilon -1},  \label{PTE} \\
&&w_{t}+ww_{x}=0\quad \rightarrow \quad u_{t}-if(u)(iu_{x})^{\varepsilon }=0.
\notag
\end{eqnarray}%
This is seen as follows: Defining $u=g(v)$ it is easy to verify that the
deformed equation in (\ref{InvB}) is converted into the deformed inviscid
Burgers equation for $v$%
\begin{equation}
v_{t}-iv(iv_{x})^{\varepsilon }=0,  \label{v}
\end{equation}%
provided the constraint%
\begin{equation}
v=f[g(v)][g^{\prime }(v)]^{\varepsilon -1}  \label{vv}
\end{equation}%
holds. Since we know already that $w\mapsto \varepsilon
v(iv_{x})^{\varepsilon -1}$ maps the inviscid Burgers equation to the
deformation (\ref{v}), we can derive from this the general map (\ref{PTE})
when using (\ref{vv}) and $v_{x}=u_{x}/g^{\prime }(v)$. Note that the
explicit form for the function $g(v)$ is only required when we want to
discuss equation (\ref{v}) in relation to the equation involving $u$.
Essential is here only its existence, which follows from the fact that (\ref%
{vv}) is separable.

Instead of $\mathcal{PT}_{\varepsilon }^{\prime }\,$, which transforms the
inviscid Burgers equation, we can also construct $\mathcal{PT}_{\varepsilon
} $, although in that case we have to be more specific about $f(w)$. Taking
for instance $f(w)=w^{n}$ the first equation in (\ref{InvB}) converts into
the second under the map%
\begin{equation}
\mathcal{PT}_{\varepsilon }:\qquad w\mapsto \sqrt[n]{\varepsilon
u(iu_{x})^{\varepsilon -1}}.  \label{wu}
\end{equation}%
Apart from the $i$, for $n=1$ this reduces to the map found by Curtright and
Fairlie \cite{Curtright:2007ta}.

These explicit maps (\ref{PTE}) and (\ref{wu}) can now be exploited to
investigate properties of the deformed equations. We start by considering in
more detail how the solutions of the $w$-equation in (\ref{InvB}) are mapped
into solutions of the deformed system. It is well known that the undeformed
equation can be solved by the method of characteristics. The characteristic,
i.e. the curve in the $xt$-plane at which $w(x,t)=w(x_{0},0)=:w_{0}(x_{0})$
is conserved, acquires in that case the form%
\begin{equation}
x=f(w_{0})t+x_{0}.  \label{ch}
\end{equation}%
The gradient catastrophe occurs when two of these characteristics cross or
equivalently when $w_{x}$ tends to inifinity. Considering therefore 
\begin{equation}
w_{x}=w_{0}^{\prime }(x_{0})\frac{dx_{0}}{dx}=\frac{w_{0}^{\prime }(x_{0})}{%
1+t\frac{df(w_{0})}{dx_{0}}},  \label{chain}
\end{equation}%
we can read off the earliest time, that is the breaking or shock time $t_{%
\text{s}}^{w}$, and the corresponding position $x_{\text{s}}^{w}$ from (\ref%
{ch}) for which this happens 
\begin{equation}
t_{\text{s}}^{w}=\min \left( -1/\frac{df(w_{0})}{dx_{0}}\right) >0\qquad 
\text{and\qquad }x_{\text{s}}^{w}=f\left[ w_{0}(x_{0}^{\min })\right] t_{%
\text{s}}^{w}+x_{0}^{\min }.  \label{ts}
\end{equation}%
Our concern here is how this translates into the deformed set of equations,
i.e. what are the corresponding times $t_{\text{s}}^{u}$ and positions $x_{%
\text{s}}^{u}$ and moreover do the deformed systems exhibit shocks?

When $f(w)=w^{n}$ the shock time resulting from (\ref{ts}) is 
\begin{equation}
t_{\text{s}}^{w}=-\frac{1}{\sqrt[n]{\varepsilon }\frac{d}{dx_{0}}\left[ \sqrt%
[n]{u_{0}}(iu_{x_{0}})^{\frac{\varepsilon -1}{n}}\right] }.  \label{tw}
\end{equation}%
For $n=1$ the expression (\ref{tw}) agrees precisely with the formula (22)
in \cite{Bender:2007ij}, derived by analysing directly the deformed equation
in (\ref{InvB}) with the more complicated method of characteristic strips.
Clearly we need to demand that the time is real, which is guaranteed when we
replace $u_{0}\rightarrow i^{\alpha }\hat{u}_{0}$ with $\hat{u}_{0}\in 
\mathbb{R}$ and $\alpha =(4m\pm 1)n/\varepsilon $, $m\in \mathbb{Z}$. Thus
for certain combinations of $\varepsilon $ and $n$ we loose the possibility
of shock wave generation for real solutions of the deformed equation.
Nonetheless, in these cases we have a correspondence between a real
undeformed solution and a complex deformed one. However, we will demonstrate
in section 4 that if we do not insist in the undeformed solution to be real
a shock formation is indeed possible, in constrast to the claims in \cite%
{Bender:2007ij}. 

Other quantities of interest which may be obtained from the explicit $%
\mathcal{PT}$-maps are conserved quantities. A conservation law for the
first equation in (\ref{InvB}) is simply derived by multiplying it with $%
\kappa f(w)^{\kappa }$ and subsequent re-arrangement 
\begin{equation}
\left[ f(w)^{\kappa }\right] _{t}+\frac{\kappa }{\kappa +1}\left[
f(w)^{\kappa +1}\right] _{x}=0.
\end{equation}%
Therefore for any asymptotically vanishing function $f(w)$ and constant $%
\kappa \in \mathbb{R}\backslash \{-1\}$ the quantities 
\begin{equation}
I_{\kappa }(w)=\int\nolimits_{-\infty }^{\infty }f[w(x,t)]^{\kappa }dx,
\label{CC}
\end{equation}%
are conserved in time. Correspondingly, we find for the deformed system the
transformed charges%
\begin{equation}
I_{\kappa }(u)=\int\nolimits_{-\infty }^{\infty }f[\varepsilon
f(u)(iu_{x})^{\varepsilon -1}]^{\kappa }dx.
\end{equation}%
We will make use of these conserved quantities below.

\section{Peak formation mechanisms from real shock waves}

Having confirmed the result for the expression of the shock times by Bender
and Feinberg in an alternative simpler manner, we diviate, however, from
their interpretation of this result. Unlike Bender and Feinberg we conclude
that these times correspond in general not to a gradient but rather to a
curvature catastrophe, i.e. the first derivative stays finite whereas the
second tends to infinity. We reason as follows: For $n=1$ it follows
obviously from (\ref{wu}) that%
\begin{equation}
w_{x}=i\varepsilon (iu_{x})^{\varepsilon -2}\left[ u_{x}^{2}+(\varepsilon
-1)uu_{xx}\right] .
\end{equation}%
This means that a possible shock in the $\mathcal{PT}$-symmetrically
deformed inviscid Burgers equation, $u_{x}\rightarrow \infty $, would always
correspond to a shock in the undeformed equation, that is $w_{x}\rightarrow
\infty $. However, the reverse does not necessarily follow as a shock in the
undeformed equation might correspond to $u_{xx}\rightarrow \infty $ with
finite $u_{x}$, rather than to $u_{x}\rightarrow \infty $. Thus in the
former scenario the shock time for the undeformed equation in (\ref{InvB})
would correspond to a formation time of a different type of wave profile in
the deformed equation in (\ref{InvB}). We can identify the explicit form by
expressing $u$ in terms of $w$ and use these expressions for our analysis.
We find%
\begin{equation}
u(x,t)=(-i)^{1-\frac{1}{\varepsilon }}(\varepsilon -1)^{\frac{1}{\varepsilon 
}-1}\varepsilon ^{\frac{\varepsilon -2}{\varepsilon }}\left[ \int^{x}w(q,t)^{%
\frac{1}{\varepsilon -1}}\,dq\right] {}^{\frac{\varepsilon -1}{\varepsilon }%
}.  \label{zz}
\end{equation}%
Differentiating this twice we obtain%
\begin{eqnarray}
u_{x}(x,t) &=&(-i)^{1-\frac{1}{\varepsilon }}(\varepsilon -1)^{\frac{1}{%
\varepsilon }}\varepsilon ^{-\frac{2}{\varepsilon }}w(x,t)^{\frac{1}{%
\varepsilon -1}}\left[ \int^{x}w(q,t)^{\frac{1}{\varepsilon -1}}\,dq\right]
{}^{-\frac{1}{\varepsilon }},  \label{zz1} \\
u_{xx}(x,t) &=&(-i)^{1-\frac{1}{\varepsilon }}(\varepsilon -1)^{\frac{1}{%
\varepsilon }-1}\varepsilon ^{\frac{\varepsilon -2}{\varepsilon }}\left(
\int^{x}w(q,t)^{\frac{1}{\varepsilon -1}}\,dq\right) {}^{-\frac{\varepsilon
+1}{\varepsilon }}w(x,t)^{\frac{2}{\varepsilon -1}}  \label{zz2} \\
&&\times \left[ \varepsilon \left( 1-\varepsilon +\int^{x}w(q,t)^{\frac{1}{%
\varepsilon -1}}\,dq\right) w_{x}(x,t)w(x,t)^{\frac{\varepsilon }{%
1-\varepsilon }}\right] ,  \notag
\end{eqnarray}%
which demonstrates that a shock in the undeformed $w$-system will lead to $%
u_{xx}\rightarrow \infty $, as it directly depends on $w_{x}$. On the other
hand $u_{x}$ only depends on $w$.

\begin{figure}[h!]
\centering   \includegraphics[width=7.5cm,height=6.0cm]{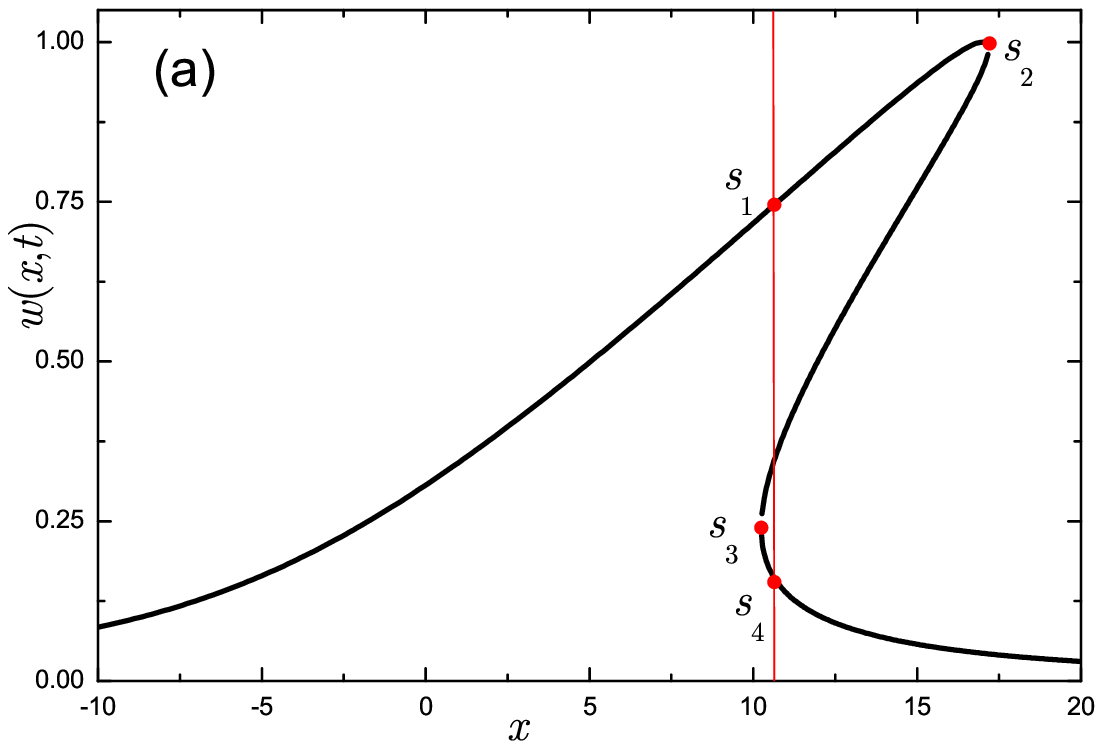} %
\includegraphics[width=7.5cm,height=6.0cm]{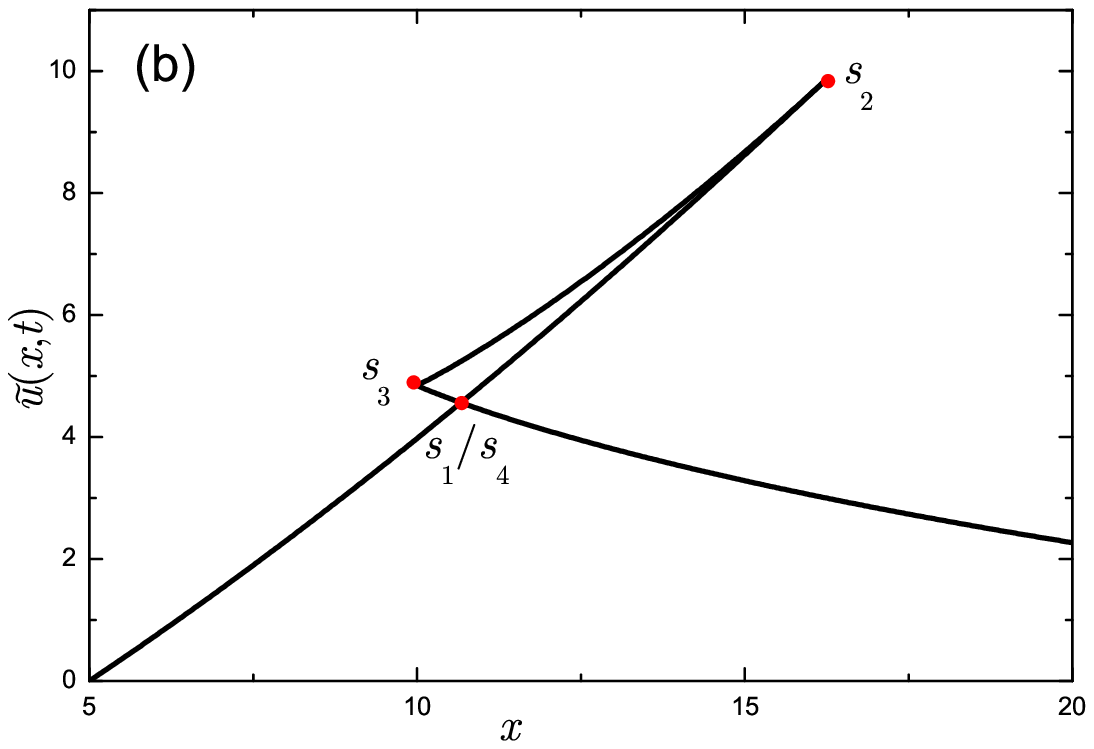}
\caption{(a) A multi-valued self-avoiding shock wave $w(x,t_{1})$. (b)
Transformed multi-valued self-crossing shock wave $\tilde{u}(x,t_{1})$ for $%
\protect\varepsilon =3$.}
\label{F1}
\end{figure}

A closer inspection of the transformation (\ref{zz}) explains how a shock is
converted into a peak by means of the $\mathcal{PT}$-deformation. Given the
form of a shock profile for real $w(x,t)$, as for instance depicted in
figure \ref{F1} panel (a), we first need to convert the multi-valued profile
into a single valued function, which is achieved by parameterising the
profile by the arc length $s$. From (\ref{zz}) we obtain%
\begin{equation}
\tilde{u}(s,t):=iu(x,t)^{\frac{\varepsilon }{\varepsilon -1}}=(\varepsilon
-1)^{-1}\varepsilon ^{\frac{\varepsilon -2}{\varepsilon -1}}\int^{s}w(q,t)^{%
\frac{1}{\varepsilon -1}}\,\frac{dq}{d\tilde{s}}d\tilde{s},  \label{trans}
\end{equation}%
with arc length element $d\tilde{s}=\sqrt{dw^{2}+dx^{2}}$. We compute $%
\tilde{u}(s,t)$ for $\varepsilon =3$, where we take the positive square root
for $s<s_{3}$ and the negative one for $s>s_{3}$. When transforming back
from $s\rightarrow x$ this function becomes multi-valued crossing itself as
depicted in figure \ref{F1} panel (b) for a time $t_{1}>t_{s}$.

This selection of the branches will produce a peaked function. In general
the choice of the different branches is naturally governed by the
appropriate boundary conditions matching the initial profile. Note that we
can eliminate the part $s_{1}\rightarrow s_{2}\rightarrow s_{3}\rightarrow
s_{4}$ without destroying the consistency of the model and thus convert a
solution from a multivalued one into a single valued peaked function. It
follows by (\ref{CC}) that the conserved quantity $I_{\kappa }$ with $\kappa
=(\varepsilon -1)^{-1}$ is unaffected by this change and remains preserved,
since for that choice $\int\nolimits_{-\infty }^{\infty
}=\int\nolimits_{-\infty }^{s_{1}}+\int\nolimits_{s_{4}}^{\infty }$.
However, for different values of $\kappa $ the $I_{\kappa }$ no longer
constitute charges for the peaked solution. This argument is similar to the
standard introduction of a shock front, the position of which is usually
chosen in such a way that $I_{1}$ is preserved. In principle this could be
implemented for the undeformed system. It is clear that changing from $%
u(x,t) $ to $\tilde{u}(s,t)$ will not alter the argument very much, apart
from introducing yet more possible branches. The peaked solutions are to be
understood in the weak sense, such that a rigorous treatment requires the
use of test functions.

We have argued above that peaked solutions are the most common ones
appearing in the deformed equations. However, under some special
circumstances, that means particular initial profiles, we may also encounter
shocks in the deformed system. We observe from (\ref{zz1}) that $%
u_{x}\rightarrow \infty $ whenever $\int^{x}w(q,t)^{\frac{1}{\varepsilon -1}%
}\,dq=0$ for $\varepsilon >0$. In turn this means by (\ref{zz}) that $%
u\rightarrow 0$. Indeed one may construct such type of solutions, as we will
demonstrate below.

\section{Jump formation from complex shocks \label{complexBurgers}}

So far we have focused on real solutions $w(x,t)$ for the inviscid Burgers
equation. However, if we wish to consider a real transformed solution $u(x,t)
$ in the case for $\varepsilon $ not being an odd integer then in the light
of (\ref{InvB}) we are forced to consider also complex initial conditions.
In fact, the complex inviscid Burgers equation has appeared before in the
literature in several different contexts, see for example \cite{BakerLi}, 
\cite{Cordoba} or \cite{Matytsin} for applications to geostrophic flows or
large N matrix models, respectively.

The method of characteristics can readily be adapted to this case. Exactly
like in the real scenario, the solution is given by $%
w(x,t)=w_{0}(x_{0}(x,t)) $, where $x_{0}(x,t)$ is found by inverting (\ref%
{ch}). Notice that in general $x_{0}(x,t)\in \mathbb{C}$ even for real $%
(x,t) $, so that the solution is defined by analytically continuing the
initial condition to complex values of the argument.\newline
In \cite{BessisFournier} it has been revealed that the shocks in inviscid
Burgers equation are due to the presence of square root singularities of $%
w(x,t)$ in the complex $x$ plane. The position of the singularities at time $%
t$ can be found by first determining $x_{0}$ by means of (\ref{chain}), and
subsequently computing $x=x_{0}-f(w_{0}(x_{0}))/\frac{df(w_{0}(x_{0}))}{%
dx_{0}}$.

For $t=0$ all such singularities are located at complex infinity.
Considering the solutions as functions of real $x$ and $t$, with $t$ being
the parameter governing the flow, the singularities move in the complex $x$%
-plane, exhibiting a shock whenever they reach the real axis. From this
discussion follows that in order to find all the possible shock times we
need to impose two conditions on $x_{0}$: 
\begin{equation}
\text{ }\func{Im}\left[ \frac{df(w_{0}(x_{0}))}{dx_{0}}\right] =0\quad \text{%
and\quad }\func{Im}\left[ x_{0}-f(w_{0}(x_{0}))/\frac{df(w_{0}(x_{0}))}{%
dx_{0}}\right] =0.  \label{conditions2}
\end{equation}

There is a crucial difference between the real and the complex case. When $%
w_{0}$, $f(w_{0})\in \mathbb{R}$, every $x_{0}\in \mathbb{R}$ solves the
second equation in (\ref{conditions2}) and we encounter a shock for all
times in some interval $[t_{s},\infty )$. The reason for this is that in the
real case the singularities reach the real axis in complex conjugate pairs
and thereafter, i.e. for $t>t_{s}$, never leave it. In contrast, in the
complex case the solutions of the second equation in (\ref{conditions2}) are
in general isolated points in the complex plane, so that a gradient
catastrophe will be an isolated event in time. The mechanism responsible for
the jump is the application of a matching condition between the intial
boundary condition and the ones for the evolved solultion. Usually the
boundary conditions are taken to be physical, that is asymptotically
vanishing. Unlike as in the real case, one can no longer stitch the two
asymptotic solutions $w_{1/2}\rightarrow \pm \infty $ together in a
continuous manner, but instead one is forced to introduce a jump.

We shall now support and illustrate our general findings with some numerical
studies.

\section{Numerical case studies}

\subsection{Real $w$ and real $u$}

The $\varepsilon =3$ deformation with $f(w)=w$ is the simplest example
allowing for real solutions for the undeformed as well as for the deformed
equation. In order to be able to compare directly with the results in \cite%
{Bender:2007ij} we consider here the same initial profile of the form of a
Cauchy distribution $u_{0}=(1+x^{2})^{-1}$, such that by (\ref{wu}) the
corresponding initial profile in the undeformed equation results to $%
w_{0}=-12x^{2}/(1+x^{2})^{5}$. According to (\ref{tw}), the gradient
catastrophe occurs therefore when 
\begin{equation}
t_{\text{gc}}^{w}(x_{0})=\frac{(1+x_{0}^{2})^{6}}{24x_{0}(4x_{0}^{3}-1)}.
\end{equation}%
This function has two distinct minima, which we identify both as shock/peak
times 
\begin{eqnarray}
x_{0,1}^{\min } &=&\frac{1}{6\sqrt{2}}\sqrt{23-\sqrt{385}}\approx 0.216621%
\text{, \ \ \ }t_{\text{s,1}}^{w}=\frac{\left( 95-\sqrt{385}\right) ^{6}}{%
2^{21}3^{10}\left( 5\sqrt{11}-2\sqrt{35}\right) }\approx 0.311791,
\label{tws} \\
x_{0,2}^{\min } &=&\frac{-1}{6\sqrt{2}}\sqrt{23+\sqrt{385}}\approx -0.769392%
\text{, \ \ \ }t_{\text{s,2}}^{w}=\frac{\left( 95+\sqrt{385}\right)
^{6}\left( 5+\sqrt{385}\right) ^{-1}}{2^{18}3^{10}\sqrt{2(23+\sqrt{385)}}}%
\approx 0.644466.  \notag
\end{eqnarray}%
In a system with shocks the second time is usually not easy to realise
numerically, but for the deformed systems these times correspond to peaks
and are directly accessible in numerical simulations. The numerical values
for the times agree with those provided in \cite{Bender:2007ij}. In addition
we compute from (\ref{ts}) the corresponding positions of the shocks/peaks as%
\begin{equation}
x_{\text{s,1}}^{w}=\frac{3\left( 19\sqrt{385}-365\right) }{64\left( 5\sqrt{11%
}-2\sqrt{35}\right) }\approx 0.0770263,\quad x_{\text{s,2}}^{w}=-\frac{3%
\sqrt{\frac{1}{2}\left( 23+\sqrt{385}\right) }}{16}\approx -1.21712.
\label{xws}
\end{equation}

\begin{figure}[h!]
\centering \includegraphics[width=7.5cm]{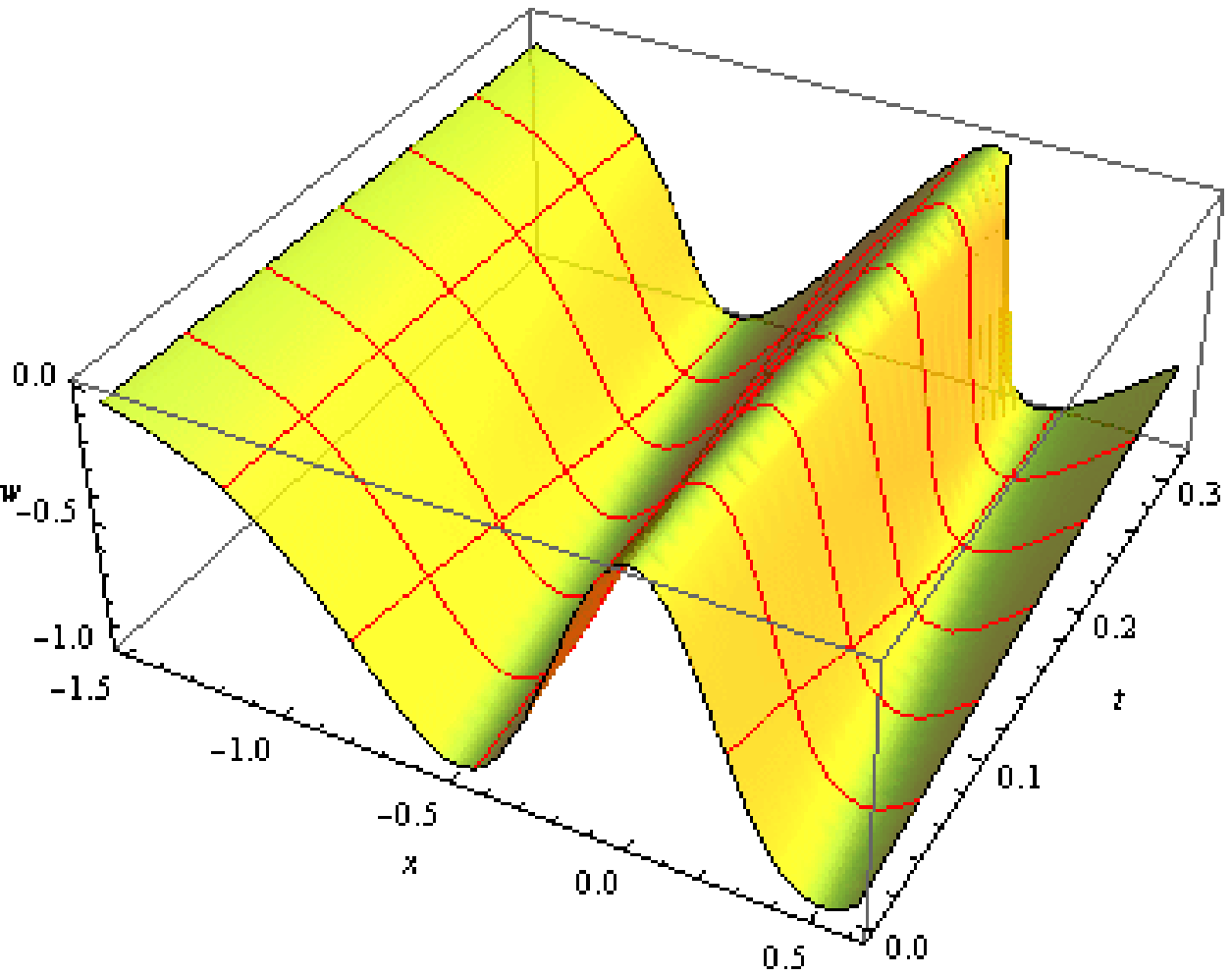} %
\includegraphics[width=7.5cm]{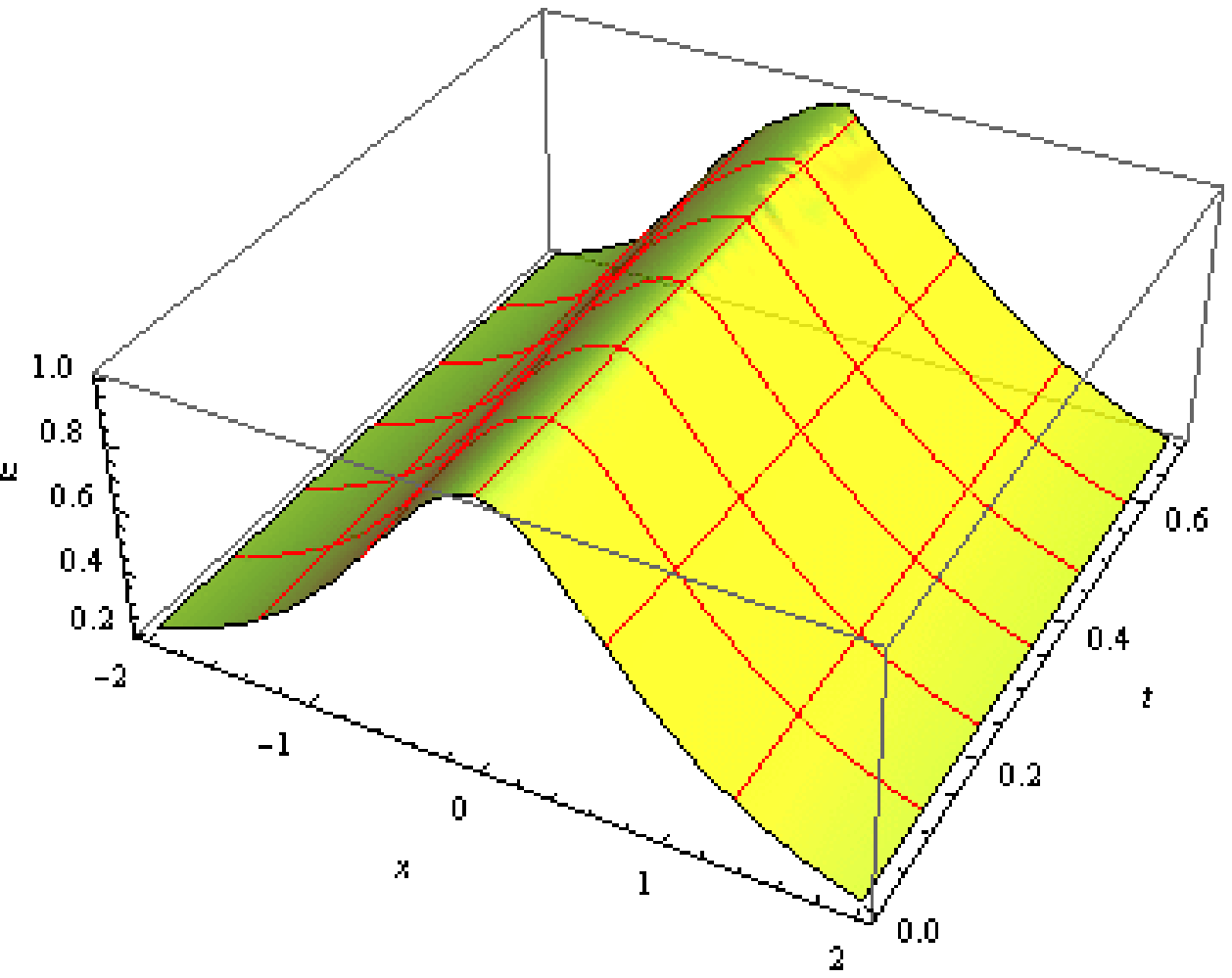} 
\caption{(a) Solution of the inviscid Burgers equation for transformed
Cauchy distribution initial profile. (b) Solutions of the $\protect%
\varepsilon =3$-deformation with Cauchy distribution initial profile.}
\label{fig1}
\end{figure}

\noindent Figure \ref{fig1} panel (a) exhibits how a shock develops in the
undeformed system at the time $t_{\text{s,1}}^{w}$ and position $x_{\text{s,1%
}}^{w}$ as predicted by (\ref{tws}) and (\ref{xws}), respectively. In panel
(b) we observe that a peak develops at the same times and positions $t_{%
\text{p,1}}^{u}=t_{\text{s,1}}^{w}$, $x_{\text{s,1}}^{u}=x_{\text{s,1}}^{w}$
and also at $t_{\text{p,2}}^{u}=t_{\text{s,2}}^{w}$, $x_{\text{s,2}}^{u}=x_{%
\text{s,2}}^{w}$. For the deformed system the numerical integration over the
discontinuities does not pose any major obstacle and the event of the second
peak can be simulated simply by integrating until that time.

In contrast, for the undeformed system this is not possible in such a
straightforward manner because after the first shock time the function $%
w(x,t)$ becomes multi-valued, such that the standard procedure becomes
meaningless. In principle this problem can be overcome by introducing a
shock front and the preservation of some conserved quantities as argued
above. For a detailed survey on these techniques see e.g. \cite{LeV}. Here
this is not needed and instead we can use the fact that our undeformed
system is implicitly solved by $w=w_{0}(x-wt)$ which we can solve
numerically for $w(x,t)$. Subsequently $u(x,t)$ is computed from (\ref{zz}).

\begin{figure}[h!]
\centering   \includegraphics[width=7.5cm,height=6.0cm]{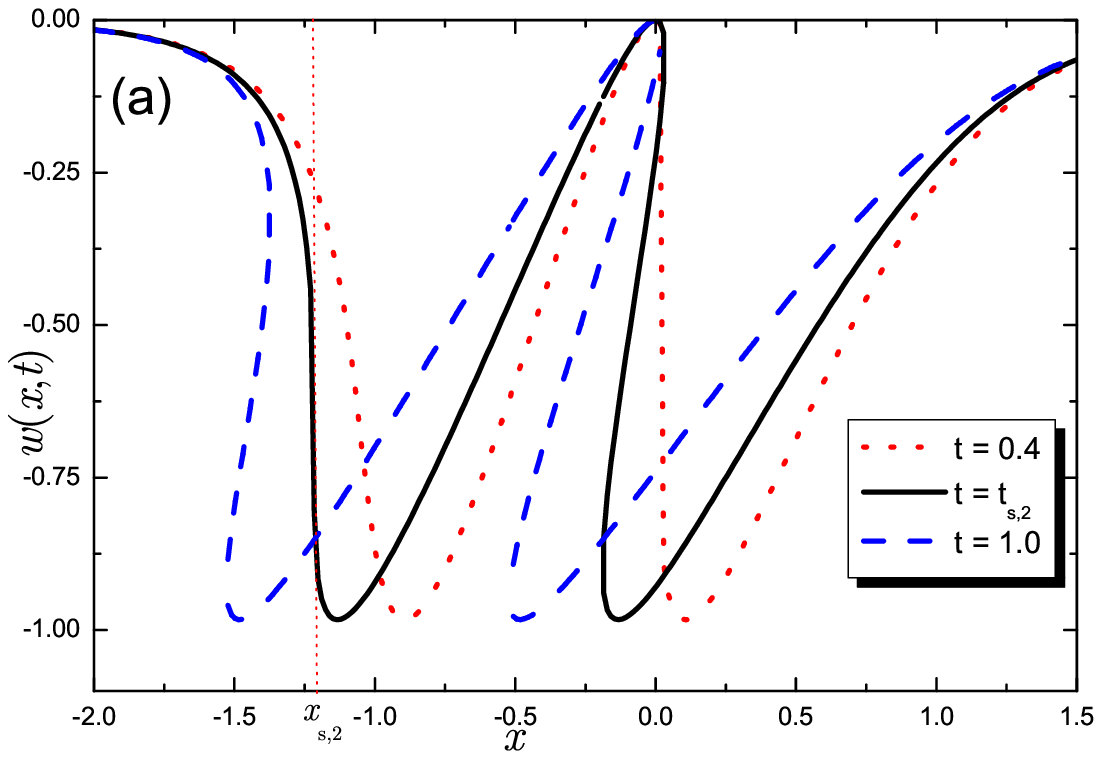} %
\includegraphics[width=7.5cm,height=6.0cm]{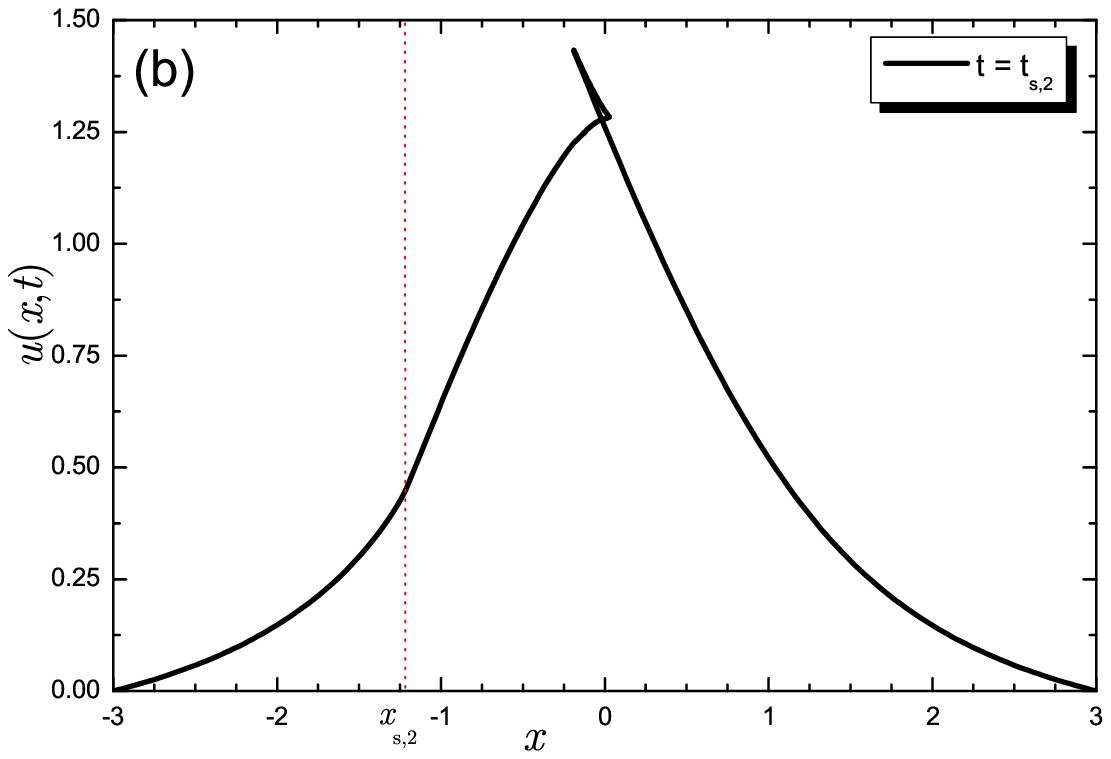}
\caption{(a) Solution of the inviscid Burgers equation at times before and
after the second shock formation with $t=0.4$ dotted (red), the second shock
time $t=t_{s,2}$ solid (black) and $t=1.0$ dashed (blue) for transformed
Cauchy distribution initial profile. (b) Solution of the $\protect%
\varepsilon =3$-deformation of the inviscid Burgers equation at the second
shock time $t=t_{s,2}$ for transformed Cauchy distribution initial profile.}
\label{fig:fronts1}
\end{figure}

The results of this computation are depicted in figure \ref{fig:fronts1},
which shows that the second shock time $t_{\text{s,2}}$ and position $x_{%
\text{s,2}}$ are approached at their predicted values (\ref{tws}) and (\ref%
{xws}), respectively. These positions in space and time coincide with those
of the second peak in the deformed system. In panel (b) we observe that the
effect on the smoothness of the curve is much less pronounced for the second
"peak". In this case it is hardly visible due to fact that it is not located
on the crest of the wave.

\begin{figure}[h]
\centering   \includegraphics[width=7.5cm,height=6.0cm]{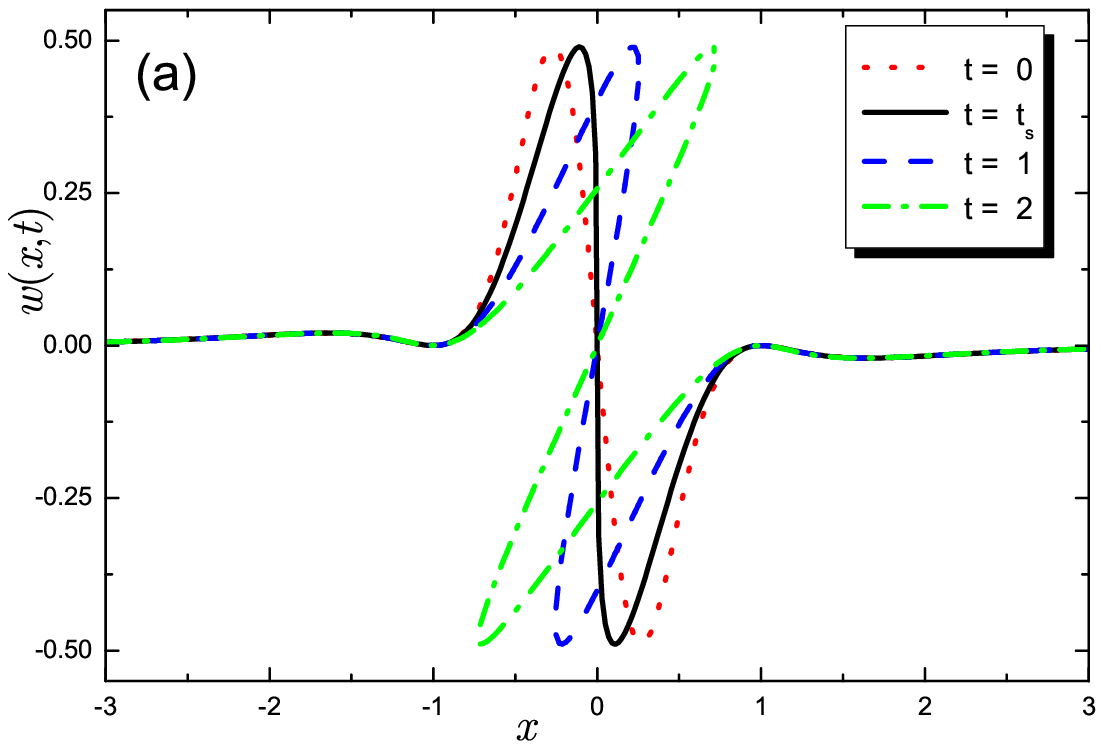} %
\includegraphics[width=7.5cm,height=6.0cm]{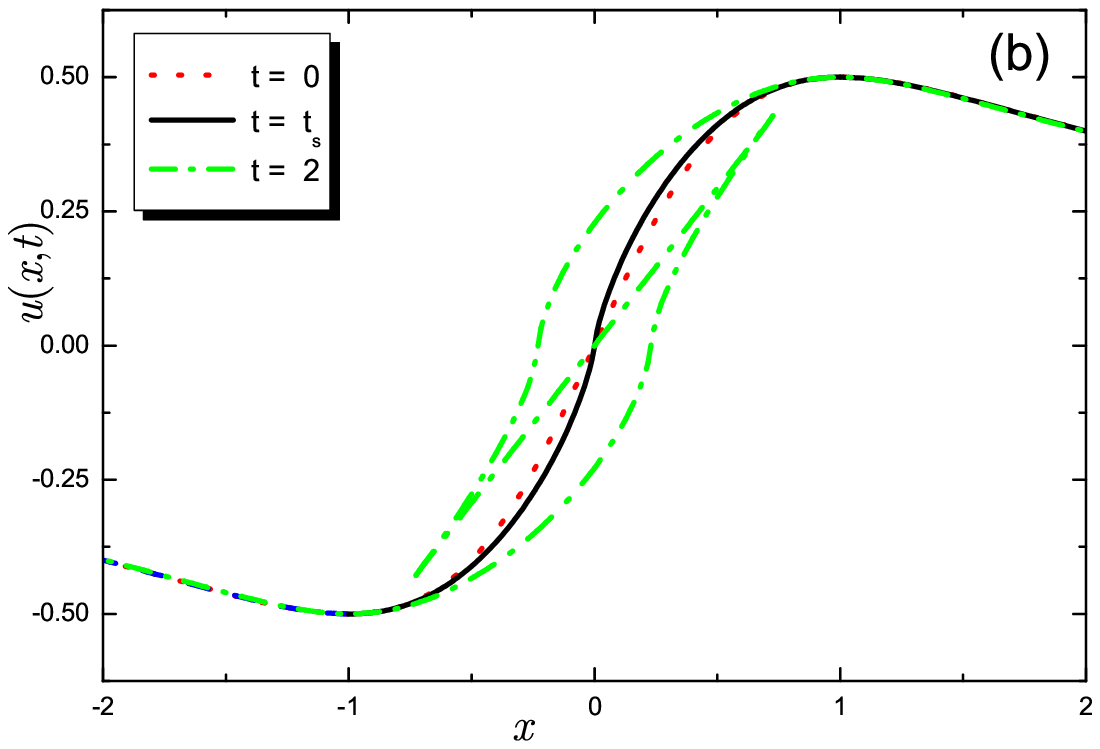}
\caption{(a) Shock wave formation at $w=0$ for the inviscid Burgers equation
at times $t=0$ dotted (red), the shock time $t_{s}=1/3$ solid (black), $t=1$
dashed (blue) and $t=2$ dasheddotted (green) for the transformed initial
profile $x/(1+x^{2})$. (b) Solution of the $\protect\varepsilon =3$%
-deformation of the inviscid Burgers equation at $u=0$.}
\label{static}
\end{figure}

Let us now present an example for the formation of a shock rtaher than a
peak. For the deformed system we take as initial profile $u_{0}=x/(1+x^{2})$%
, such that the initial profile of undeformed equation results to $%
w_{0}=-3x(1-x^{2})^{2}/(1+x^{2})^{5}$. According to (\ref{ts}) this system
develops a shock at time $t_{\text{s}}^{w}=1/3$ at position $x_{\text{s}%
}^{w}=0$. For the undeformed system these findings are clearly confirmed by
our numerical results depicted in figure \ref{static} panel (a). Notice in
panel (b) that when the solution evolves beyond the shock time two more
shocks, in the sense that $u_{x}\rightarrow \infty $, develop when $u$
becomes zero. This behaviour is forced by the relation between $w$ and $u$, (%
\ref{PTE}), since $w$ is finite.

\subsection{Real $w$ and complex $u$}

The $\varepsilon =2$ deformation with $f(w)=w$ is the easiest case to
investigate the scenario for which $w\in \mathbb{R}$ and $u\notin \mathbb{R}$%
. We compute initially the time and position of the shock and peaks in the
undeformed system (\ref{InvB}), respectively.

\begin{figure}[h!]
\centering  \includegraphics[width=7.5cm]{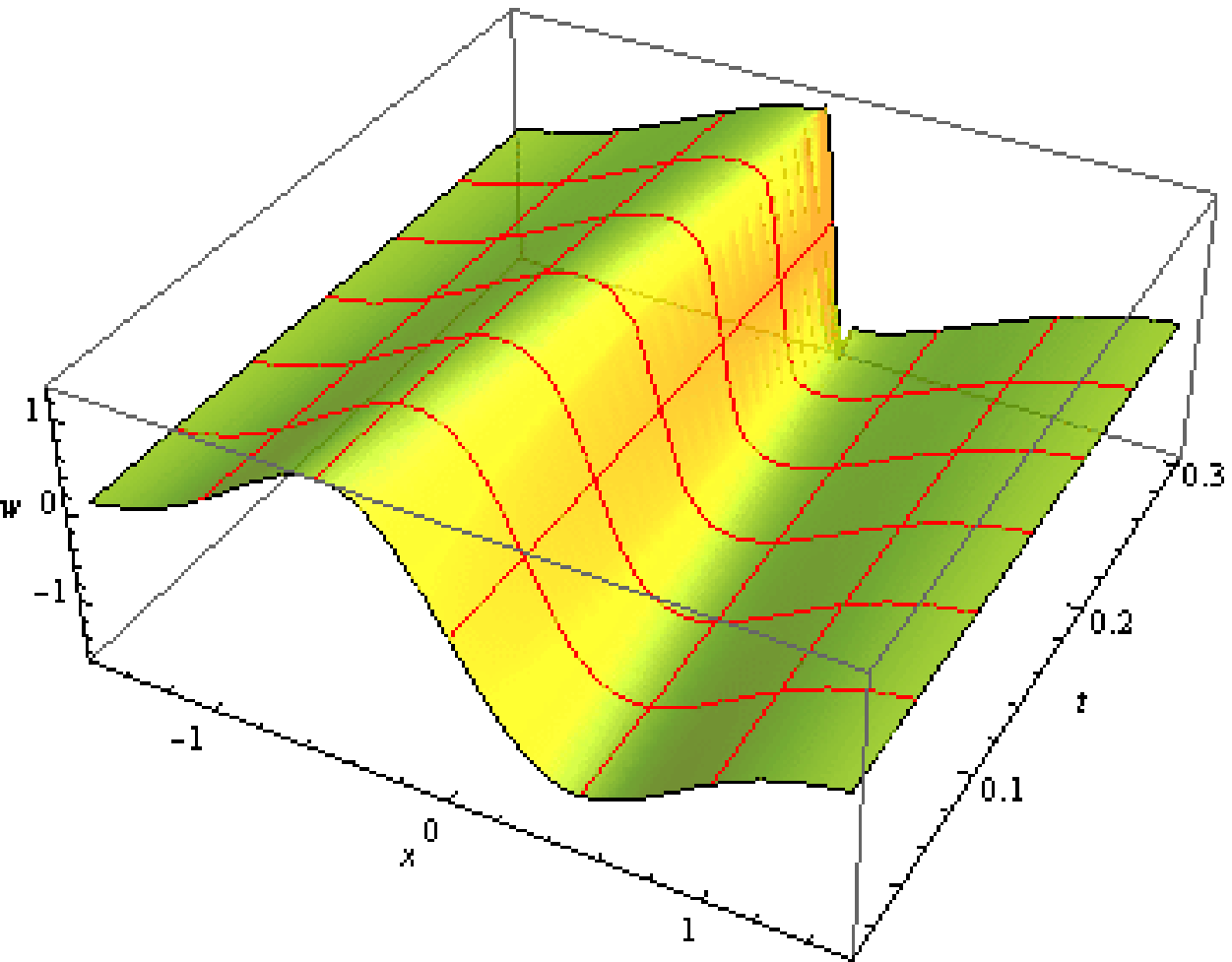} %
\includegraphics[width=7.5cm]{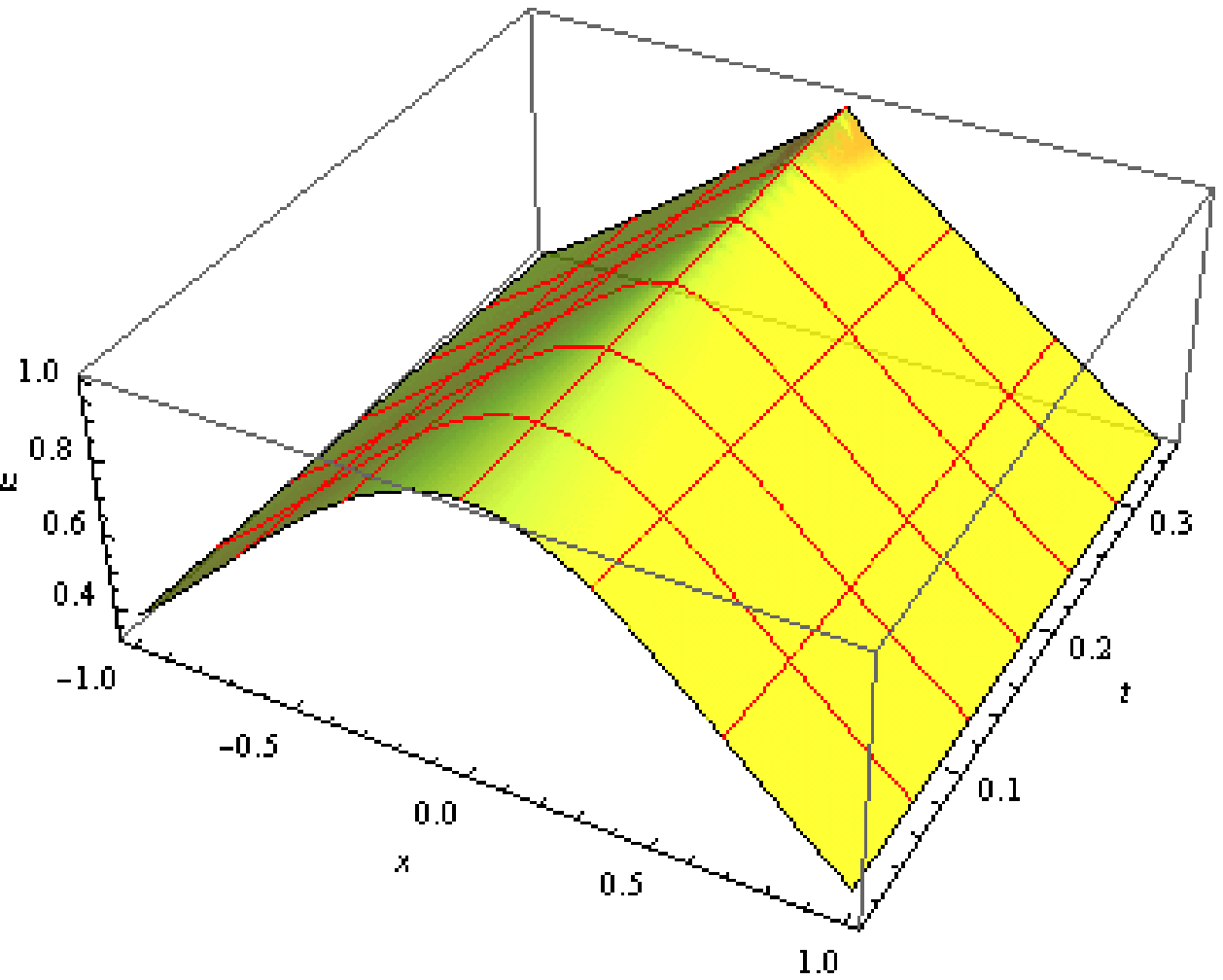} 
\caption{(a) Solutions of the inviscid Burgers equation for a transformed
Gaussian initial profile. (b) Solution of the $\protect\varepsilon =2$%
-deformation with a Gaussian initial profile in the $\exp (i\protect\pi /4)$%
-direction.}
\label{fig11}
\end{figure}

We take the initial profile to be of Gaussian form $u_{0}=e^{-x^{2}-i\frac{%
\pi }{4}}$ leading to $w_{0}=-4xe^{-2x^{2}}$ with the help of the
transformation (\ref{wu}). As argued above, to take the phase in $u_{0}$ is
one possibility to guarantee real shock times. The time for the gradient
catastrophe computed from (\ref{tw}) is $t_{\text{gc}}^{w}=$ $%
e^{-x_{0}^{2}}/(4-16x^{2})$, which becomes minimal for\ $x_{0}^{\min }=0$.
The resulting shock/peak time and position are therefore $t_{\text{s}%
}^{w}=t_{\text{s}}^{u}=1/4$ and $x_{\text{s}}^{w}=x_{\text{s}}^{u}=0$,
respectively.

Figure \ref{fig11} panel (a) clearly shows how a shock develops in the
undeformed system at the predicted shock time $t_{\text{s}}$ and position $%
x_{\text{s}}$. Panel (b) exhibits that in the deformed system this shock is
converted into a peak occurring at the same time and position. In general
this wave travels in the complex $u$-plane, but here we have only plotted
here the $\exp (i\pi /4)$-direction for which $u(x,t)$ becomes real. We also
note that the peak becomes more pronounced as the wave evolves beyond its
time of formation.

\subsection{Complex $w$ and complex $u$}

Here we will consider the deformation with $\varepsilon =3/2$ . As discussed
in section \ref{complexBurgers}, we now require a complex solution for the
undeformed equation in order to generate shock waves. We select a simple
initial condition which vanishes asymptotically 
\begin{equation}
w_{0}(x)=w(x,0)=\frac{e^{\frac{i\pi }{4}}}{x^{2}+1}.  \label{eq:w_init}
\end{equation}%
and in addition allows to solve the well known implicit realtion $%
w=w_{0}(x-wt)$. The transformation (\ref{PTE}) then guarantees that we have
real initial conditions for the deformed equation 
\begin{equation}
u_{0}(x)=\left[ \frac{4}{3}\int_{-\infty }^{x}\frac{1}{(y^{2}+1)^{2}}\,dy%
\right] ^{1/3}{},  \label{eq:u_init}
\end{equation}%
with boundary conditions 
\begin{equation}
\lim\limits_{x\rightarrow -\infty }u_{0}(x)=0,\quad
\lim\limits_{x\rightarrow +\infty }u_{0}(x)=:k\approx 1.2794,\quad \text{%
and\quad }\lim\limits_{x\rightarrow \pm \infty }\partial _{x}u_{0}(x)=0.%
\text{ }  \label{bu}
\end{equation}%
We demand that these boundary conditions are preserved for $t>0$.

Given $w_{0}$, we compute the shock time as outlined in section \ref%
{complexBurgers}. The conditions (\ref{conditions2}) have two solutions: $%
z_{0,1}\approx 0.164903-0.553299i$ and $z_{0,2}=-z_{0,1}$, and,
correspondingly, we find two shock times: $t_{s,1}\approx 0.4791>0$ and $%
t_{s,2}=-t_{s,1}<0$. The shock occurring for positive time is located at $%
x_{s,1}\approx 0.494709$.

\begin{figure}[h!]
\centering  \includegraphics[width=7.5cm]{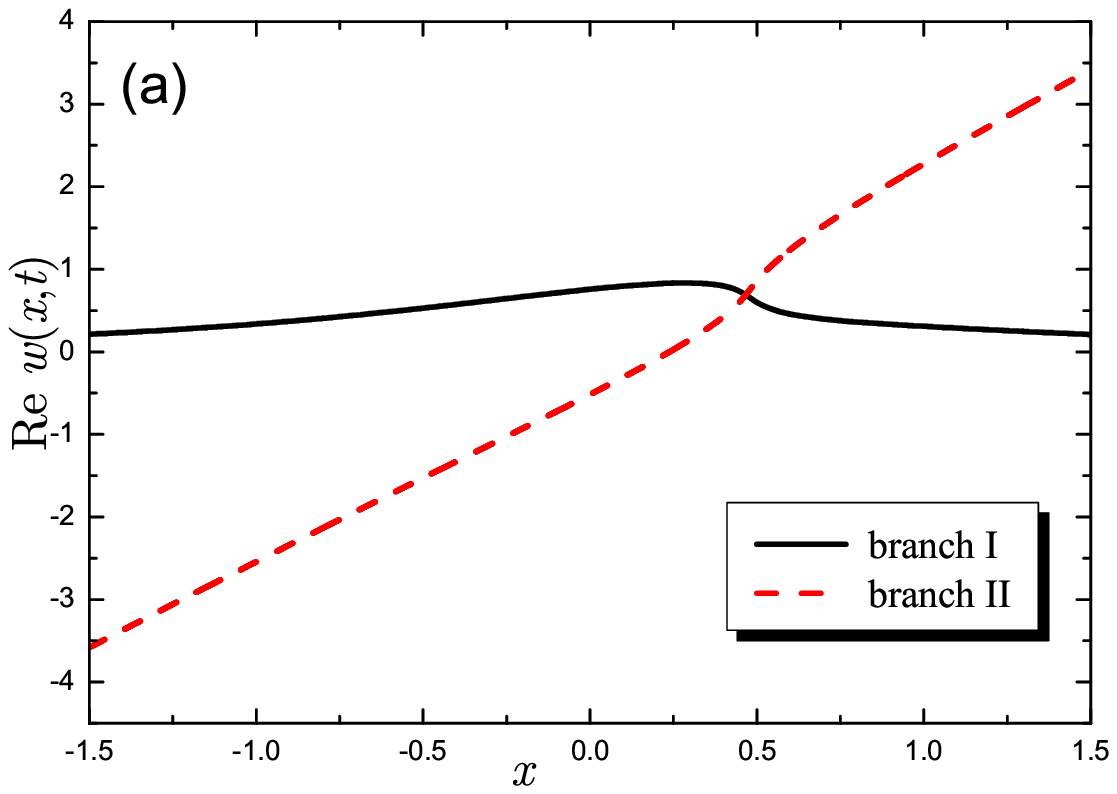} %
\includegraphics[width=7.5cm]{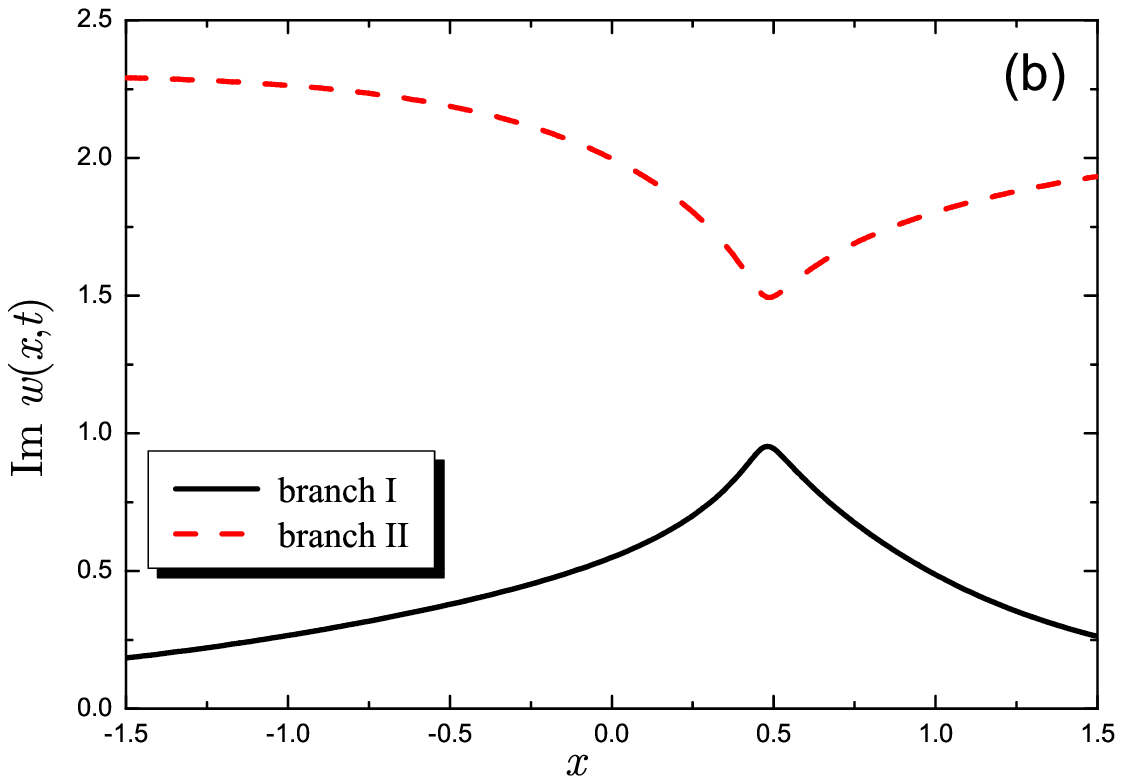} %
\includegraphics[width=7.5cm]{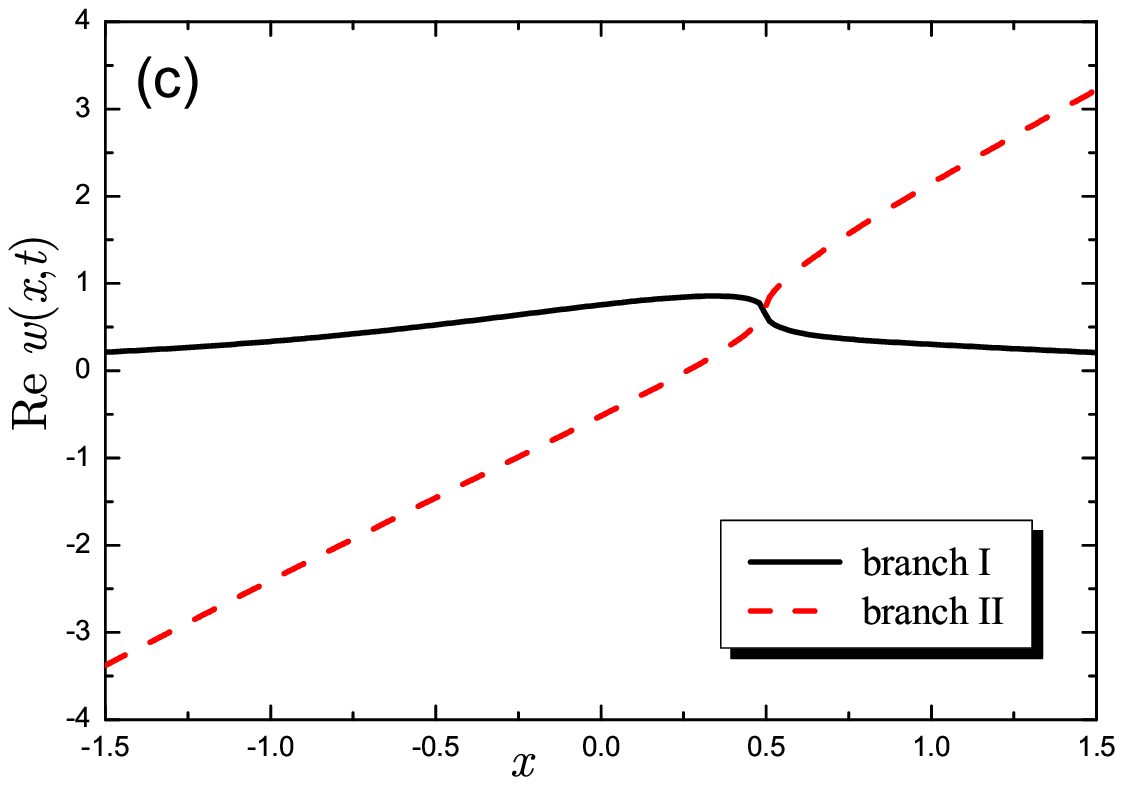} %
\includegraphics[width=7.5cm]{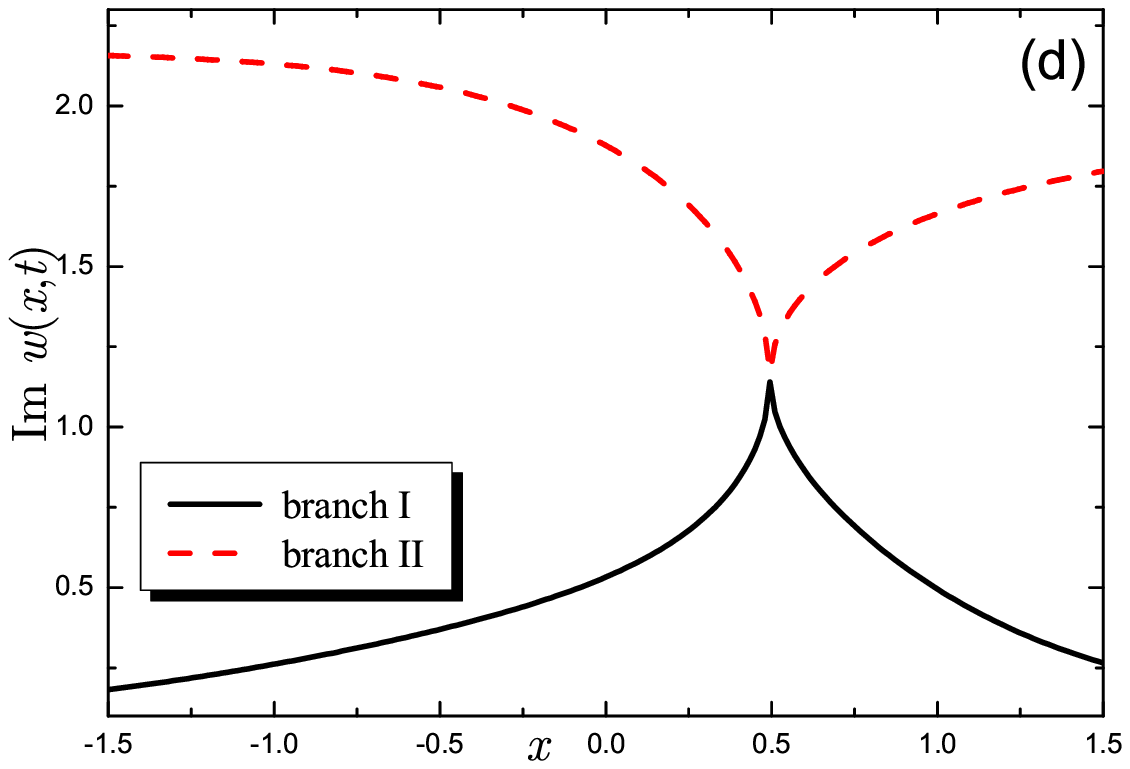} %
\includegraphics[width=7.5cm]{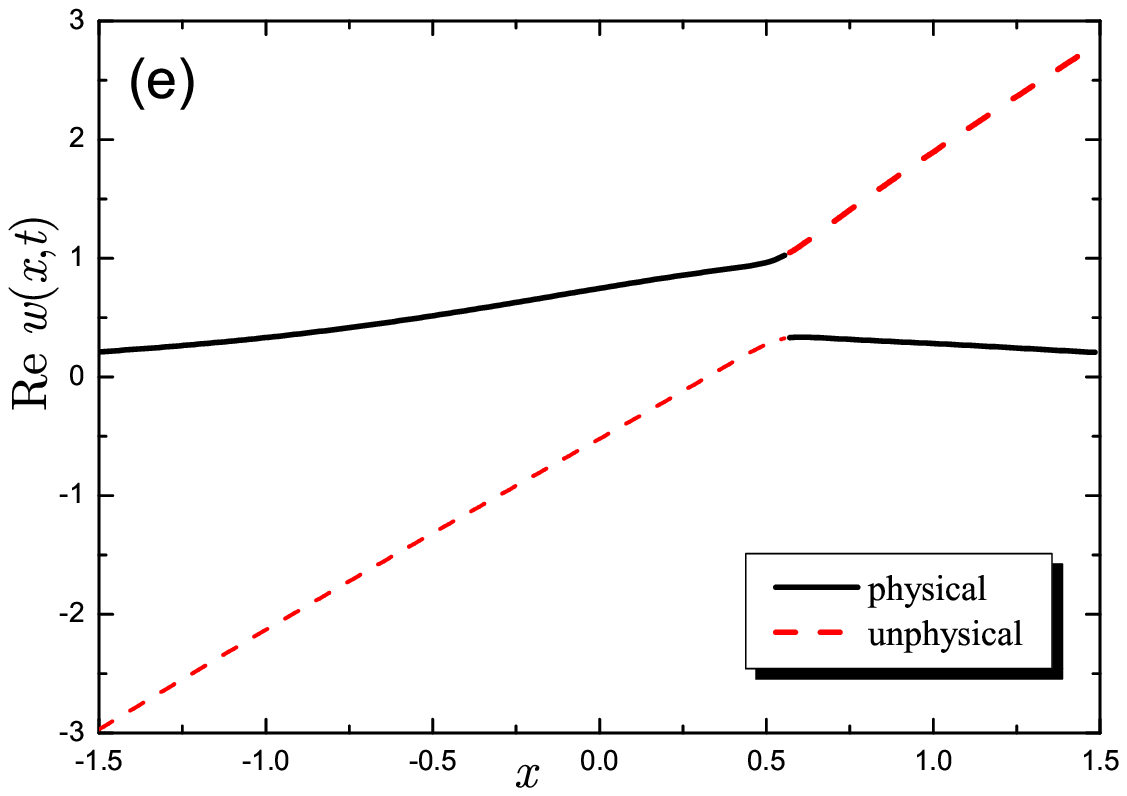} %
\includegraphics[width=7.5cm]{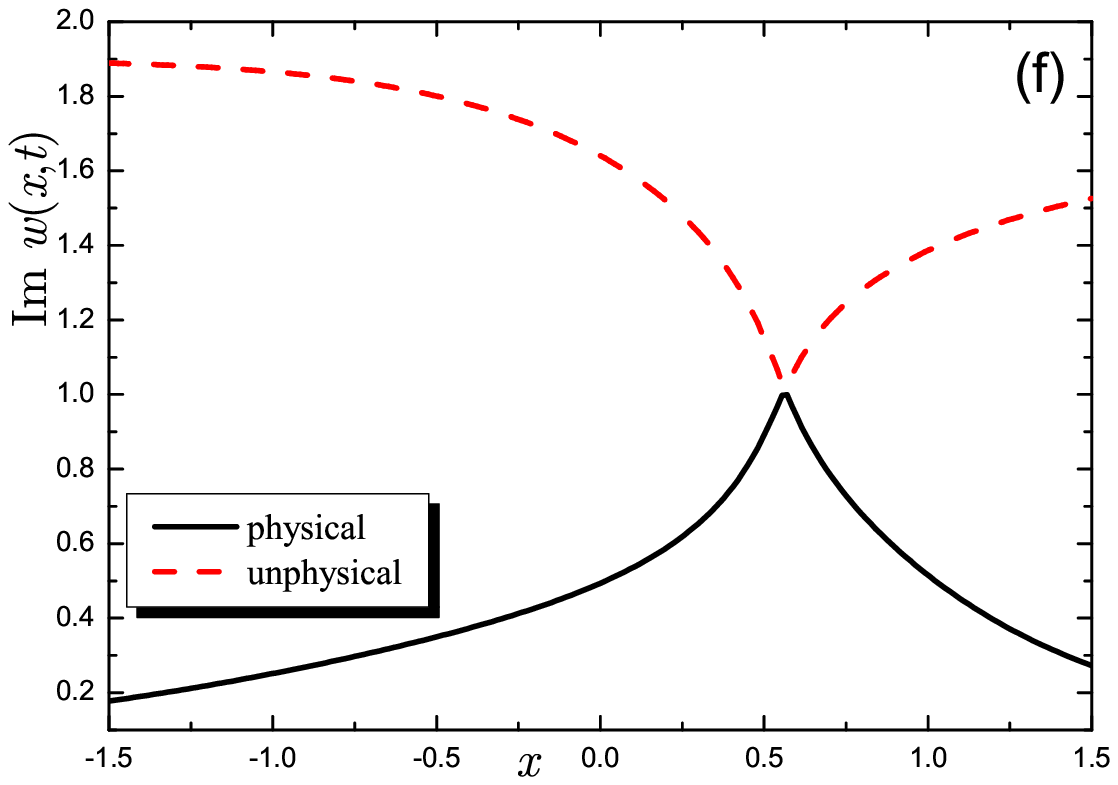} \centering  
\caption{ (a), (b) Real and imaginary part of the solution of the inviscid
Burgers equation with complex initial condition (\protect\ref{eq:w_init})
solid (black) and a second, unphysical branch dashed (red) at time $t = 0.45
< t_{s, 1}$. (c), (d) The two branches touching at the shock time $t = t_{s,
1}$. (e), (f) The physical solution with vanishing asymptotic boundary
conditions exhibits a jump discontinuity at $t = 0.55 > t_{s, 1}$. }
\label{wcomplex1}
\end{figure}

In figure \ref{wcomplex1}, we plot real and imaginary part of the solution $%
w(x,t)$, obtained by inverting the relation $w=w_{0}(x-wt)$. In this
example, $w(x,t)$ has three branches, of which only two are represented in
the figure. The branch represented as a solid (black) line is the solution
satisfying the initial condition (\ref{eq:w_init}).

We see that the two branches touch at $t=t_{s,1}$. For $t>t_{s,1}$, the left
part of one branch has connected with the right part of the other, imposing
a jump on the physical solution in order to preserve the asymptotic
behaviour $\lim\nolimits_{x\rightarrow \pm \infty }w(x,t)=0$.

\begin{figure}[h!]
\centering   \includegraphics[width=7.5cm]{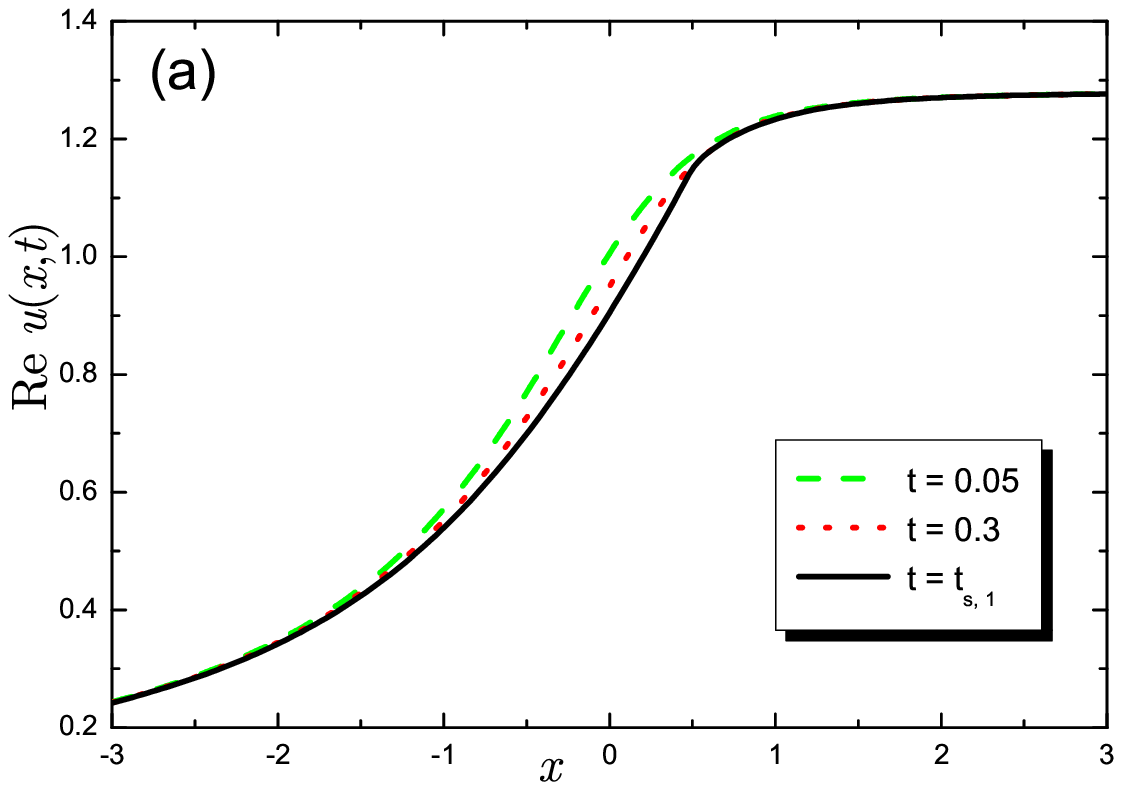} %
\includegraphics[width=7.5cm]{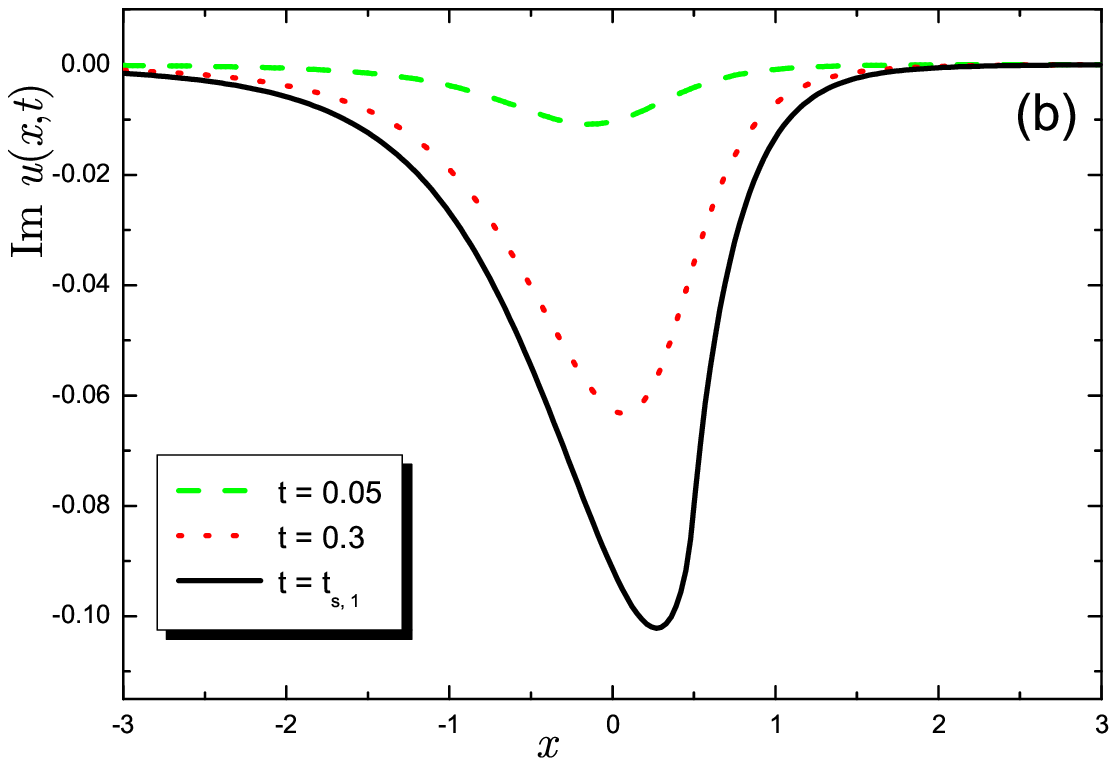}
\caption{Real and imaginary parts of the solution of the deformed Burgers
equation with $\protect\varepsilon = 3/2$ and initial condition (\protect\ref%
{eq:u_init}) for $t = 0.05$ dashed (green), $t= 0.3$ dotted (red) and at the
shock time $t = t_{s, 1}$ solid (black).}
\label{ucomplex1}
\end{figure}

To see how this reflects on the evolution of the deformed field, we have
constructed the solution $u(x,t)$ using the relation (\ref{zz}) 
\begin{equation}
u(x,t)=\left[ -\frac{4i}{3}\int_{-\infty }^{x}w(y,t)^{2}\,dy\right] ^{1/3}{},
\label{eq:left_limit}
\end{equation}%
and evaluated the integral numerically. In figure \ref{ucomplex1}, we
represent $u(x,t)$ for a sequence of times leading to $t_{s,1}$, and we
observe again that $u_{x}$ is continuous, while $u_{xx}(x_{s,1})\rightarrow
\infty $ as $t\rightarrow t_{s,1}$.\newline

After the shock time, $w(x,t)$ develops a jump. We will show that the
boundary conditions (\ref{bu}) impose a jump on $u(x,t)$ as well. In order
to see this we will try to match the following expressions: 
\begin{eqnarray}
\hat{u}(x,t) &=&\left( -\frac{4i}{3}\int_{-\infty
}^{x}w^{(1)}(y,t)^{2}\,dy\right) {}^{\frac{1}{3}};  \label{eq:limits} \\
\tilde{u}(x,t) &=&\left( k^{3}-\frac{4i}{3}\int_{+\infty
}^{x}w^{(2)}(y,t)^{2}\,dy\right) ^{\frac{1}{3}},
\end{eqnarray}%
where the two branches $w^{(1,2)}(x,t)$ are defined by: $w^{(1)}(-\infty
,t)=0=w^{(2)}(+\infty ,t)$ for $t>t_{s,1}$. By construction, $\hat{u}(x,t)$
satisfies the left boundary condition in (\ref{bu}), while $\tilde{u}(x,t)$
satisfies the condition on the right.

\begin{figure}[h!]
\centering   \includegraphics[width=7.5cm]{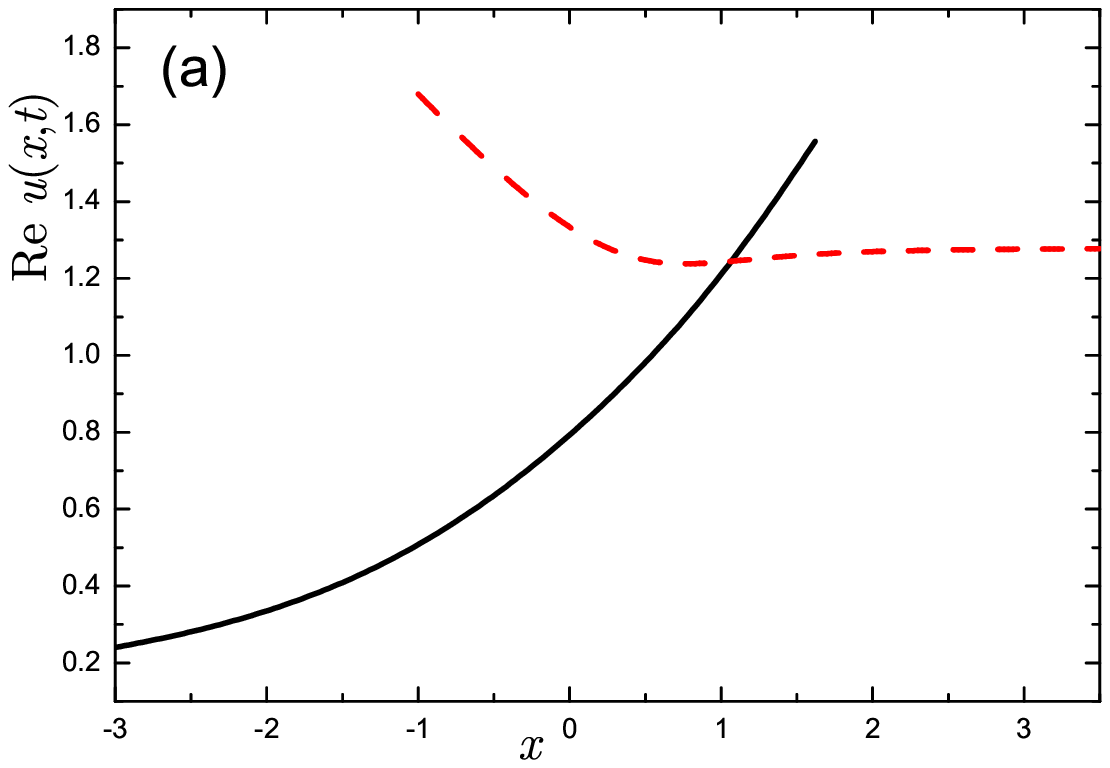} %
\includegraphics[width=7.5cm]{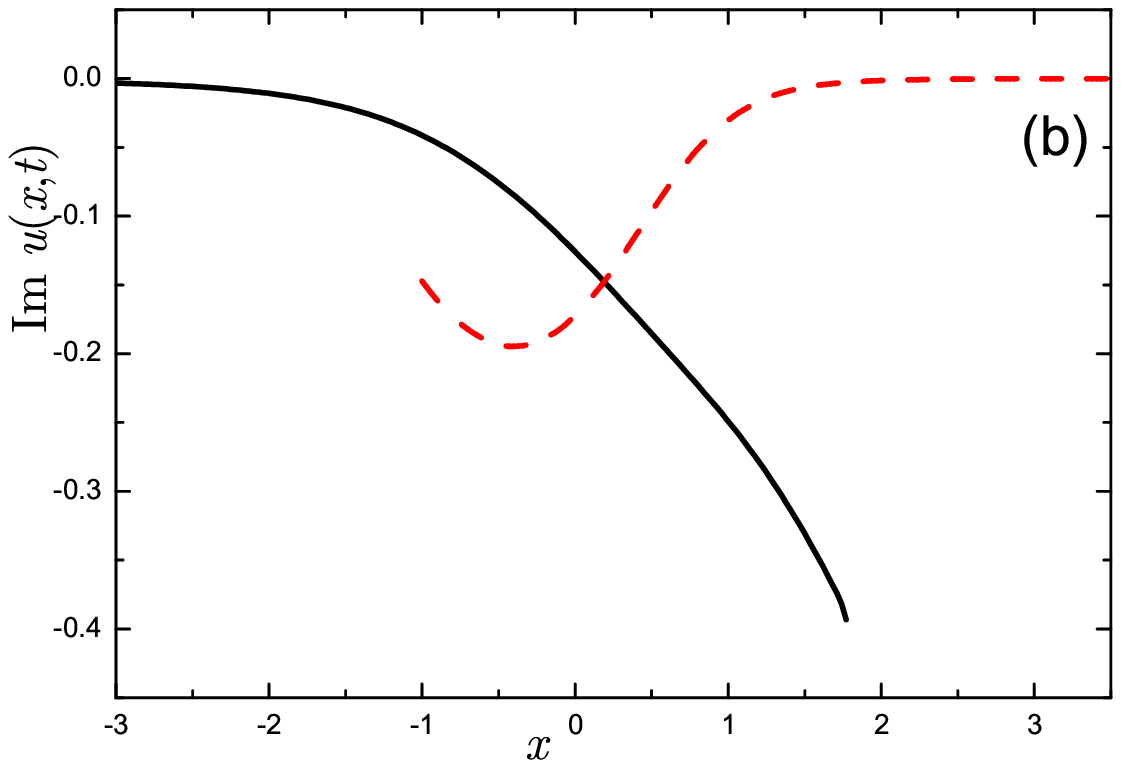}
\caption{Real and imaginary parts of the two functions $\hat{u}(x, t)$ solid
(black) and $\tilde{u}(x, t)$ dotted (red) defined in (\protect\ref%
{eq:limits}), for $t = 1.$ The condition $\text{Re } \hat{u}(x, t) = \text{%
Re }\tilde{u}(x, t)$ is satisfied for $x = x_1 \approx 1.0663$, and the
condition $\text{Im } \hat{u}(x, t) = \text{Im }\tilde{u}(x, t)$ for $x =
x_2 \approx 0.1893 \neq x_1$.}
\label{ucomplex2}
\end{figure}

In figure \ref{ucomplex2} we show $\hat{u}(x,t)$ and $\tilde{u}(x,t)$ for $%
t=1.$ We notice that, contrary to the real case treated in section 3, we can
not find an $x_{\ast }$ such that $\hat{u}(x_{\ast },t)=\tilde{u}(x_{\ast
},t)$. This can be understood because the former condition now splits into
two real equations, while we have only one real parameter $x_{\ast }$ to
tune. The consequence is that a continuous solution for $u(x,t)$ does not
exist beyond $t_{s,1}$.

\subsection{Deformations with odd $\protect\varepsilon $ and $f(w)=w$}

We will now study how the systems behave as a function of increasing values
of the deformation parameter $\varepsilon $. We keep the initial profile in
the deformed system to be a Cauchy distribution $u_{0}=(1+x^{2})^{-1}$, such
that by (\ref{wu}) the initial profile in the undeformed equation will
change. We do not report the explicit expressions for these functions here,
but only the resulting shock and peak times the following table:

\begin{center}
\begin{tabular}{|c||c|c|c|c|}
$\varepsilon $ & $t_{\text{s,1}}$ & $t_{\text{s,2}}$ & $x_{\text{s,1}}$ & $%
x_{\text{s,2}}$ \\ \hline\hline
3 & 0.311791 & 0.644466 & 0.0770262 & -1.21712 \\ \hline
5 & 0.394011 & 0.662872 & -0.18255 & 1.05226 \\ \hline
7 & 0.594697 & 0.913866 & 0.241058 & -0.970114 \\ \hline
9 & 0.997223 & 1.45053 & -0.279227 & 0.919109 \\ \hline
11 & 1.78617 & 2.50127 & 0.306641 & -0.883621 \\ \hline
13 & 3.34619 & 4.555 & -0.327569 & 0.857142%
\end{tabular}
\end{center}

\begin{figure}[h!]
\centering   \includegraphics[width=7.5cm]{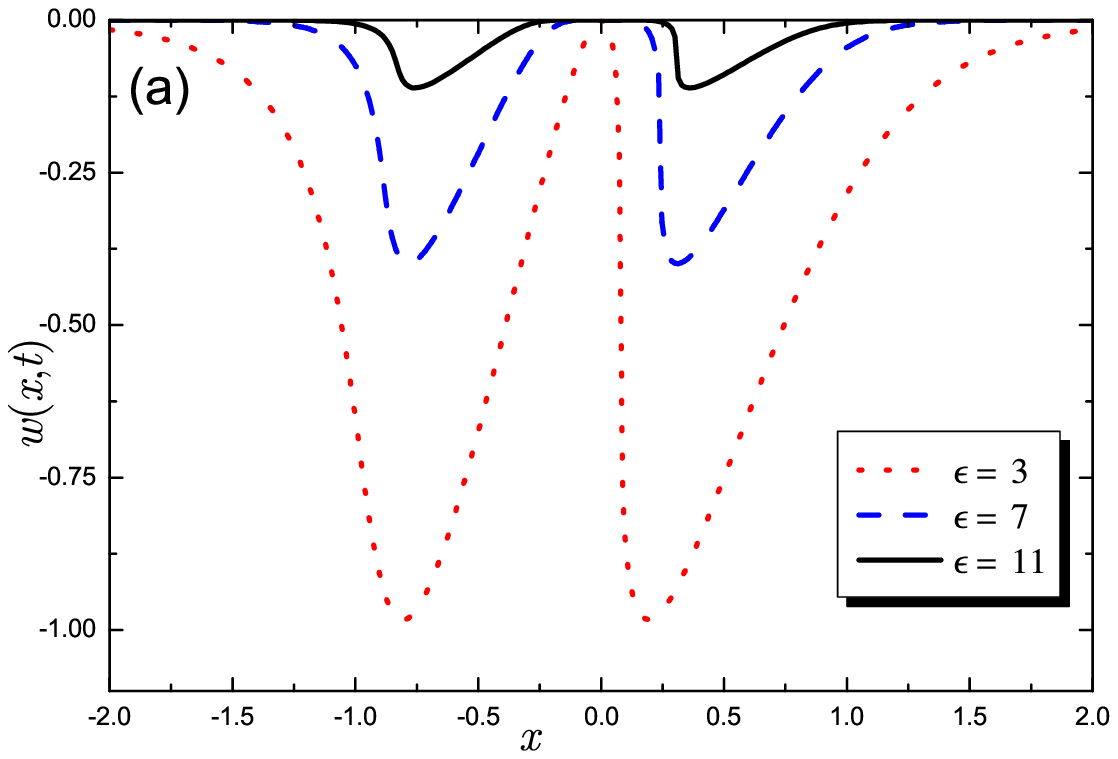} %
\includegraphics[width=7.5cm]{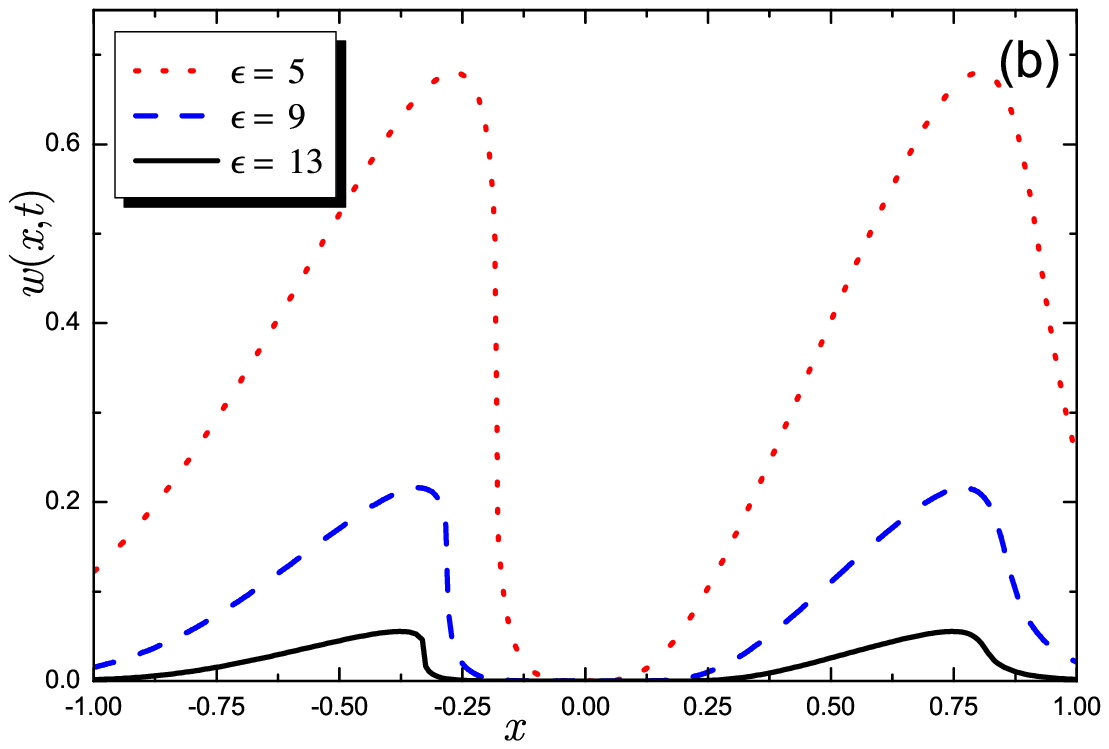} %
\includegraphics[width=7.5cm]{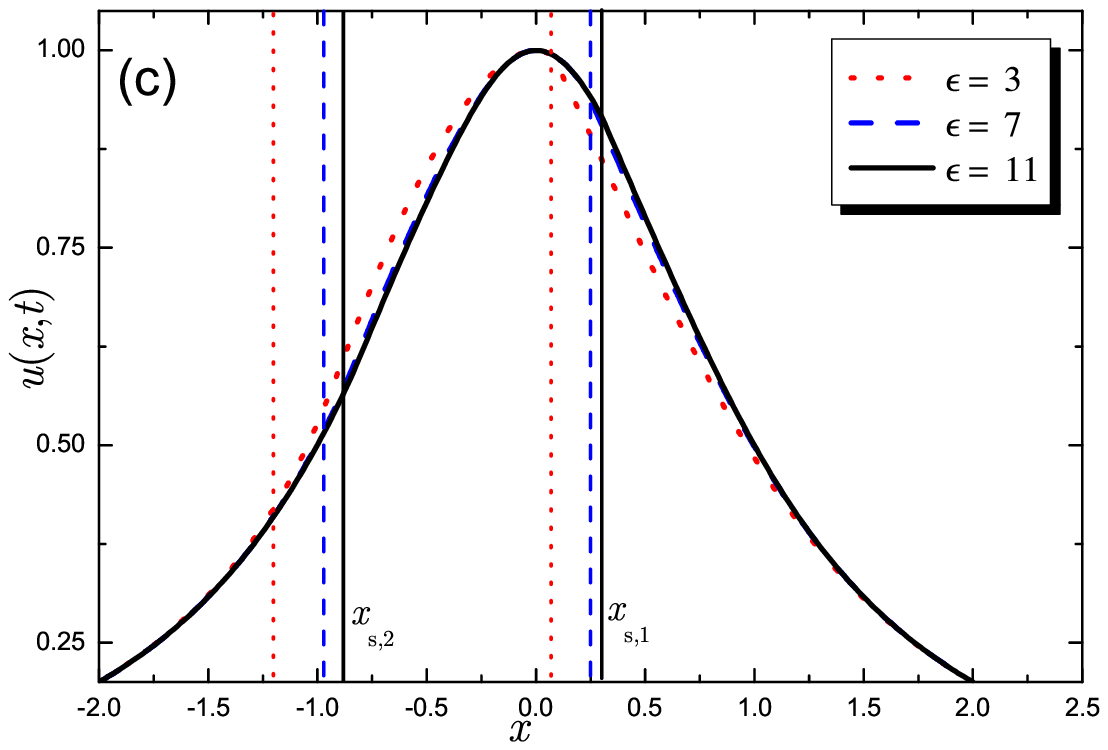} %
\includegraphics[width=7.5cm]{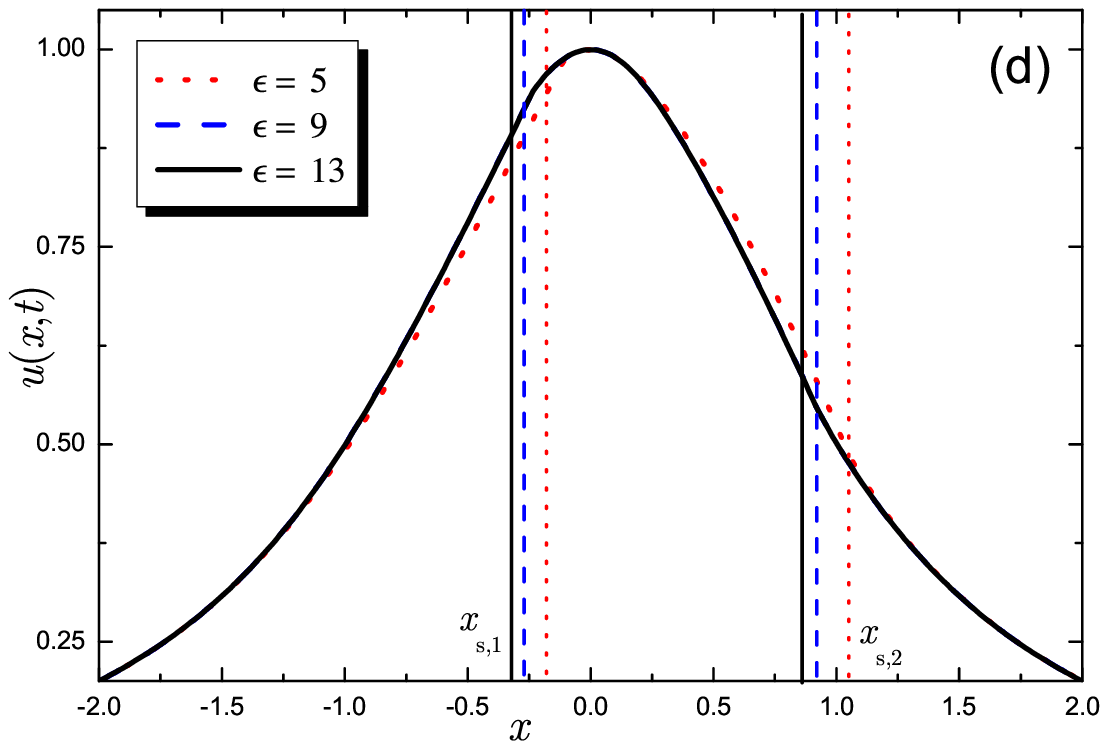} 
\caption{(a), (b) Solutions of the undeformed system at the shock time $t_{%
\text{s,1}}$ for various values of $\protect\varepsilon$ with transformed
Cauchy distribution initial profile. (c), (d) Solution of the deformed
system at the shock time $t_{\text{s,1}}$ for a Cauchy distribution initial
profile.}
\label{fig4abc}
\end{figure}

In figure \ref{fig4abc} panels (a), (b) we observe that the first shock
times are reproduced in our numerical solutions and that they occur at the
predicted locations.

\noindent We observe that the overall qualitative behaviour does not change
with increasing values of $\varepsilon $, but the amplitutes are
considerably reduced. We also note that the systems with $\varepsilon =4m-1$
and $\varepsilon =4m+1$ with $m\in \mathbb{N}$ break first on the right and
left wave front, respectively. In the panels (c) and (d) we observe that the
shocks have been smoothed out considerably and that the larger $\varepsilon $
becomes the less pronounced they are. The peaks are extremely small when
they occur on the tails of the waves rather than on its crest.

\subsection{Multi-peak solutions}

\begin{figure}[h]
\centering  \includegraphics[width=7.5cm]{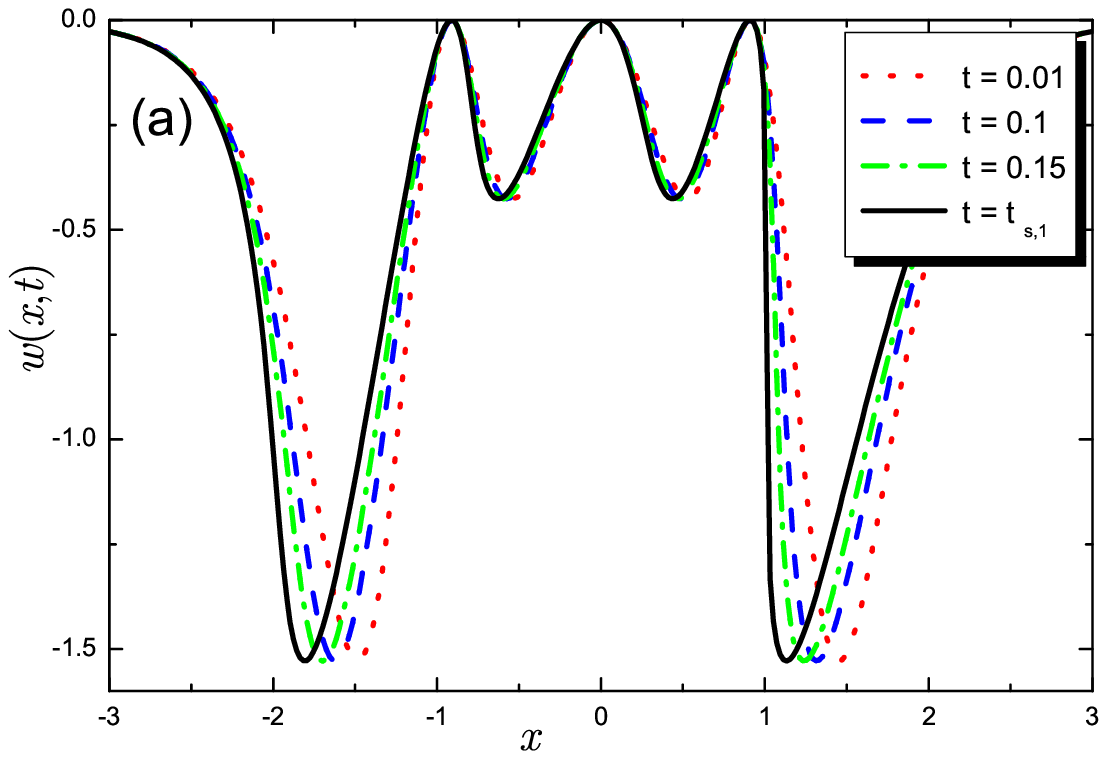} %
\includegraphics[width=7.5cm]{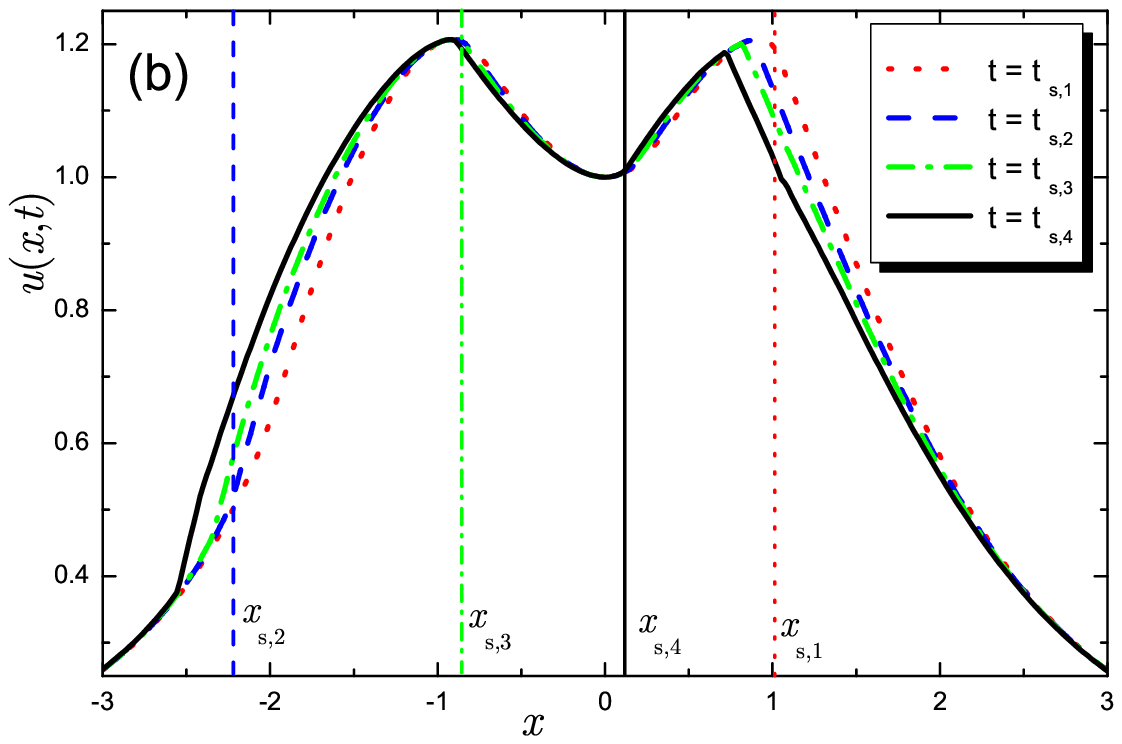}
\caption{(a) Solutions of the inviscid Burgers equation for transformed sum
of shifted Cauchy distribution initial profile at shock peak times $t_{\text{%
s1}}$ dotted (red), $t_{\text{s2}}$ dashed (blue), $t_{\text{s3}}$
dasheddotted (green) and $t_{\text{s4}}$ solid (black). (b) Solutions for
its $\protect\varepsilon =3$-deformation at the same times and Cauchy
distribution initial profile.}
\label{fig5}
\end{figure}

In the same way as for as for peakons an interesting question is whether it
is possible to obtain multi-peaked solutions. We demonstrate here that the
answer to this is affirmative. As an example we consider the equations with $%
f(w)=w$ and deformation parameter $\varepsilon =3$. Taking then for instance
the initial profile in the deformed system to be a sum of two shifted Cauchy
distributions $u_{0}=[1+(x-1)^{2}]^{-1}+[1+(x+1)^{2}]^{-1}$, the initial
profile of undeformed equation results to $w_{0}=-96\left( x^{2}+2\right)
\left( x^{5}+4x^{3}-4x\right) ^{2}\left( x^{4}+4\right) ^{-5}$. By means of (%
\ref{tw}) the time for the gradient catastrophe is then computed to 
\begin{equation}
t_{\text{gc}}^{w}(x_{0})=-\frac{\left( x^{4}+4\right) ^{6}}{384x\left(
2x^{14}+25x^{12}+60x^{10}-156x^{8}-384x^{6}+240x^{4}+192x^{2}-64\right) }.
\end{equation}%
Minimising this function we find four shock/peak times $t_{\text{s1}%
}=0.221045$, $t_{\text{s2}}=0.429609$, $t_{\text{s3}}=0.558845$, $t_{\text{s4%
}}=0.798264$ and corresponding positions $x_{\text{s1}}=1.01299$, $x_{\text{%
s2}}=-2.21359$, $x_{\text{s3}}=-0.856069$, $x_{\text{s4}}=0.116185$. All
these values are accurately reproduced in figure \ref{fig5}. Once again the
peaks emerging on the crest of the wave are well pronounced.

\section{Conclusions}

For large a class of nonlinear wave equations of inviscid Burgers type (\ref%
{InvB}) we have shown that their real shock wave solutions are smoothed out
in their $\mathcal{PT}$-symmetrically deformed counterparts into peaks. The
mechanism for the peak formation was identified to be the folding of a
self-avoiding multi-valued function into self-crossing multi-valued
function. Under the preservation of certain conservation laws one can
consistently eliminate the looping part of the self-crossing multi-valued
function and thus converting it into a single valued weak solution of the
deformed wave equation.\ In general, we found that the larger the
deformation parameter $\varepsilon $ the smoother the peaks. Our analytical
arguments were facilitated by the explicit knowledge of the $\mathcal{PT}$%
-transformation from one system to the other.

We also showed that shocks in the complex solutions for the undeformed
equations will lead to discontinuities in the solutions for the deformed
equations. 

It would be of great interest to construct more explicit maps for different
types of nonlinear wave equations to their $\mathcal{PT}$-symmetrically
deformed counterparts, such as for instance for the KdV equation. 

As a practical application one may potentially use these observation in
numerical investions of nonlinear wave equations with shock wave formations.
In general those equations are difficult to handle, even numerically, but
one can exploit our obervations by first solving the deformed equations,
which are simpler to deal with as they just exhibit peaks and then transform
those solutions to the undeformed system.

\medskip

\noindent \textbf{Acknowledgments:} AC is supported by a City University
Research Fellowship.


\end{document}